\documentclass{ws-ijmpd}
\usepackage[super,compress]{cite}
\usepackage{booktabs}
\usepackage{pifont}
\usepackage{color}
\usepackage[colorlinks]{hyperref}

\definecolor{CiteColor}{rgb}{0,0.5,0}
\hypersetup{citecolor=CiteColor}
\definecolor{RefColor}{rgb}{0.55,0,0}
\hypersetup{linkcolor=RefColor}

\newcommand{\beq}{\begin{equation}}
\newcommand{\eeq}{\end{equation}}
\newcommand{\ud}{\mathrm{d}}
\newcommand{\ui}{\mathrm{i}}
\newcommand{\calF}{\mathcal{F}}
\newcommand{\calO}{\mathcal{O}}

\begin{document}

\markboth{Alexandre Le Tiec}{The overlap of numerical relativity, perturbation theory and post-Newtonian theory\dots}

%\catchline{}{}{}{}{}
%
%%%%%%%%%%%%%%%%%%%%%%%%%%%%%%%%%%%%%%%%%%%%%%%%%%%%%%%%%%%%%%%%%%%%

\title{The Overlap of Numerical Relativity, Perturbation Theory and Post-Newtonian Theory in the Binary Black Hole Problem}

\author{Alexandre Le Tiec}

\address{Laboratoire Univers et Th\'eories (LUTh), Observatoire de Paris, CNRS, \\ Universit\'e Paris Diderot, 5 place Jules Janssen, 92190 Meudon, France \\ letiec@obspm.fr}

\maketitle

%\begin{history}
%\received{Day Month Year}
%\revised{Day Month Year}
%\end{history}

\begin{abstract}
Inspiralling and coalescing binary black holes are promising sources of gravitational radiation. The orbital motion and gravitational-wave emission of such system can be modelled using a variety of approximation schemes and numerical methods in general relativity: the post-Newtonian formalism, black hole perturbation theory, numerical relativity simulations, and the effective one-body model. We review recent work at the multiple interfaces of these analytical and numerical techniques, emphasizing the use of coordinate-invariant relationships to perform meaningful comparisons. Such comparisons provide independent checks of the validity of the various calculations, they inform the development of a universal, semi-analytical model of the binary dynamics and gravitational-wave emission, and they help to delineate the respective domains of validity of each approximation method. For instance, several recent comparisons suggest that perturbation theory may find applications in a broader range of physical problems than previously thought, including the radiative inspiral of intermediate mass-ratio and comparable-mass black hole binaries.
\end{abstract}

\keywords{Black holes; gravitational waves; numerical relativity; post-Newtonian theory; perturbation theory.}

\ccode{PACS numbers: 04.25.-g, 04.25.D-, 04.25.dg, 04.25.Nx, 04.30.-w}

%\tableofcontents

\section{Introduction}

The ``two-body problem'' has always played a central role in gravitational physics.\cite{Da.87} In Einstein's general theory of relativity, the simplest and most universal two-body problem is that of a binary system of black holes. The coalescence of two black holes produces copious amounts of gravitational radiation,\cite{Pr.09} whose observation would have a tremendous impact on physics, astrophysics and cosmology \cite{SaSc.09}. A worldwide network of ground-based, kilometer-scale laser interferometric gravitational-wave antennas (LIGO, Virgo, GEO600, KAGRA) is currently under development, with the goal of detecting gravitational radiation from inspiralling compact-object binaries by the end of this decade.\cite{Aa.al.13} Moreover, space-based gravitational-wave detectors such as the proposed eLISA mission will observe the coalescence of two supermassive black holes in distant galaxies, as well as the inspiral of stellar mass compact bodies into massive black holes.\cite{Am.al.13} If black holes with masses in the range $\sim 10^2-10^4 M_\odot$ exist in globular clusters, intermediate mass-ratio inspirals would be another highly promising source of gravitational waves for both ground-based and space-based detectors \cite{Mi2.09,AmSa.10}.

\subsection{Modelling binary black holes}

Nevertheless, the promise of gravitational-wave astronomy crucially relies upon our ability to accurately model the gravitational-wave signals generated by such astrophysical sources. Indeed, the detection and analysis of these signals require highly accurate theoretical predictions, for use as \textit{template waveforms} to be cross-correlated against the output of the detectors by means of matched filtering.\cite{Ba.al.13} The orbital dynamics and gravitational-wave emission of binary black holes can be modelled using a variety of analytical approximation schemes and numerical techniques within general relativity. Below we provide a brief overview of existing methods and refer the reader to more topical reviews for additional information.

\vspace{-0.3cm}

\subsubsection*{Post-Newtonian theory} 

The post-Newtonian (PN) approximation to general relativity is the natural tool to model inspiralling black hole binaries with small orbital velocities/large separations, but otherwise arbitrary mass ratios. In PN theory, general relativistic corrections to the Newtonian solution are incorporated into the equations of motion and the radiation field order by order in the small parameter $v^2 \sim M / r \ll 1$, where $v$ and $r$ are the typical orbital velocity and separation,\footnote{Throughout this paper we set $G = c = 1$ and use a metric signature $+2$. Latin indices $a,b,\dots$ are abstract, while the letters $i,j,\dots$ refer to spatial components in a particular frame.} and $M \equiv m_1 + m_2$ the sum of the component masses. Several approaches to the problem of motion have been developed to a high degree of sophistication, such as (i) the PN iteration of the Einstein equation in harmonic coordinates,\cite{Bl.al.98,BlFa.01,deA.al.01,Bl.al.04,PaWi.02,MiWi.07} (ii) the canonical Hamiltonian formalism in ADM-TT coordinates,\cite{JaSc.98,Da.al2.00,Da.al.01,JaSc.13,Da.al.14} (iii) a surface integral approach pioneered by Einstein, Infeld and Hoffman, \cite{It.al.01,ItFu.03,It.04} and (iv) the application of effective field theory methods.\cite{GoRo.06,FoSt.11,GaLe.12,FoSt.13} As a result, the PN equations of motion of two spinless black holes are known up to 4PN order.\footnote{The $n$PN order refers either to terms $\calO(1/c^{2n})$ in the equations of motion, with respect to the Newtonian acceleration, or in the radiation field, relatively to the standard quadrupolar waveform.} The problem of radiation has been extensively investigated within the multipolar post-Minkowskian wave generation formalism of Blanchet and Damour,\cite{BlDa.86,Bl.87,Bl.98} as well as using the ``direct integration of the relaxed Einstein equation'' approach of Will, Wiseman and Pati.\cite{WiWi.96,PaWi.00} For non-spinning binaries moving along quasi-circular (resp. quasi-eccentric) orbits, the gravitational-wave phase evolution has been computed up to 3.5PN (resp. 3PN) order,\cite{Bl.al.02,Bl.al2.02,Bl.al2.04,Bl.al2.05} while amplitude corrections are known up to 3PN order.\cite{Bl.al.96,Ar.al.04,Ki.al.07,Ki.08,Bl.al.08} Since astrophysical black holes may carry significant spins, much effort is currently being devoted to the inclusion of spin effects in PN template waveforms.\cite{St.al.08,St.al3.08,PoRo.08,PoRo2.08,He.al.10,Le.10,Le2.10,Po.10,Po.al2.11,HaSt.11,HaSt2.11,Le2.12,Po.al.12,Ha.al.13,Bl.al2.11,Bu.al.13,Ma.al.13,Bo.al.13,Bo.al2.13,Ma.al.14,LeSt.14} See Refs.~\refcite{Bl.14,Sc.11,FuIt.07,FoSt.14} and \refcite{Bl.11} for recent reviews.

\vspace{-0.3cm}

\subsubsection*{Black hole perturbation theory} Black hole perturbation (BHP) theory is the natural tool to model compact binaries with mass ratios $q \equiv m_1 / m_2 \ll 1$, even for strong-field orbits with $r \gtrsim M$ or $v \lesssim 1$. At zeroth order, the small body moves along a timelike geodesic of the background spacetime $\mathring{g}_{ab}$ of a Schwarzschild or Kerr black hole of mass $m_2$. At first order, the body's motion is affected by the metric perturbation $h_{ab} = \calO(q)$ sourced by its small mass $m_1$. The motion can be described as accelerated with respect to the background metric $\mathring{g}_{ab}$ under the effects of a gravitational self-force (GSF),\cite{Mi.al.97,QuWa.97,GrWa.08} or equivalently as geodesic in the effective metric $\mathring{g}_{ab} + h^R_{ab}$, where the regular metric perturbation $h^R_{ab}$ is a certain smooth vacuum solution of the linearized Einstein equation.\cite{DeWh.03,Po2.10,Ha.12} The self-force can be split into (i) a dissipative component whose average piece is directly related to gravitational-wave emission and (ii) a conservative component responsible for secular effects.\cite{HiFl.08} While the leading gravitational-wave emission, as described by the Regge-Wheeler-Zerilli equations in a Schwarzschild background\cite{ReWh.57,Ze3.70} and by the Teukolsky equation in Kerr,\cite{Te.72,Te.73} has been explored extensively,\cite{Sh.94,PoSa.95,Mi.al.96,Ta.al3.96,Mi.03,Hu.00,Hu.al.05,Sa.al.06} recent work has mostly focused on computing conservative effects.\cite{De.08,Sa.al.08,BaSa.09,Bl.al.10,Bl.al2.10,Ba.al.10,Sh.al.11,BaSa.11,Sh.al.14,Do.al.14,Do.al2.14}
Significant effort is currently underway to compute such conservative GSF effects for non-circular orbits\cite{BaSa.10,Wa.al.12,Wa.al.14} and Kerr backgrounds,\cite{Sh.al.12,He.al.14,Is.al.14} as well as the second-order gravitational self-force.\cite{De.12,Gr.12,Po.12,Po2.12,BuKa.13,PoMi.14,Po.14} See Refs.~\refcite{SaTa.03,Ba.09,De.11,Wa.11,Th.11} and \refcite{Po.al.11} for recent reviews.

\vspace{-0.3cm}

\subsubsection*{Numerical relativity} In general, the description of the nonlinear radiative dynamics of a binary black hole system entails a full numerical-relativistic treatment. Solving the exact vacuum Einstein equation numerically by means of supercomputer simulations is a formidable task which required overcoming numerous technical challenges over several decades (\textit{e.g.} formulations, gauge conditions, stable evolutions, black hole excision, boundary conditions, wave extraction). Following a celebrated breakthrough in 2005 that, for the first time, allowed tracking the orbital motion and extracting the gravitational radiation emitted during the last orbits, final plunge, merger and ringdown,\cite{Pr.05,Ca.al.06,Ba.al.06} the field of numerical relativity (NR) has entered a ``golden age.'' Indeed, NR simulations have made possible a systematic exploration of binary black hole spacetimes in the most nonlinear, fully relativistic regime,\cite{Ow.al.11} from the simplest case of equal-mass and non-spinning binaries,\cite{Ba.al2.06,Ba.al.07,Ba.al3.07,Bo.al2.07,Bu.al.07,Ha.al.07,Sc.al.09} to unequal-mass\cite{Be.al.07,Go.al.09,Sp.al.11,Bu.al.12} and spinning binaries with aligned spins,\cite{Be.al.08,Ha.al2.08,Ch.al.09,Ha.al.10} to generic precessing systems,\cite{Ca.al.09,Sz.al.09} including many types of orbits.\cite{Sp.al.08,Sp.al2.08,GoBr.10,Hi.al.10,Sp.al2.11,GoBr.13,Da.al2.14} Ongoing work focuses on improving the numerical accuracy of the waveforms,\cite{Zl.al.12,Hi.al2.13} generating more efficiently a large number of them,\cite{Mr.al.13} interpretating waveforms for precessing binaries,\cite{Ha.al.14,Ha.14} constructing more physically realistic initial data,\cite{ReTi.12,Ch.14} and exploring regions of the parameter space that have remained uncharted so far.\cite{LoZl.11,Lo.al.12,LoZl.13} See Refs.~\refcite{Ce.al.10,Hi.10,McW.11,Pf.12,Sp.al.13} and \refcite{Sz.14} for recent reviews.

\vspace{-0.3cm}

\subsubsection*{Effective one-body model} The idea of the effective one-body (EOB) model\cite{BuDa.99,BuDa.00} is to map the ``real'' dynamics of a compact binary system of masses $m_1$ and $m_2$ onto some (non-geodesic) effective dynamics of a reduced mass $\mu \equiv m_1 m_2 / M$ moving in the effective metric of a deformed Schwarzschild black hole of mass $M = m_1 + m_2$. By construction, the EOB Hamiltonian reproduces the known PN dynamics in the weak-field/small-velocity limit, as well as the geodesic motion of a test particle in a Schwarzschild background in the extreme mass-ratio limit. More recently, fully relativistic information coming from GSF theory has been used to inform the EOB Hamiltonian.\cite{Da.10,Ba.al.10,Ba.al.12,Ak.al.12} The initial model was also extended to account for spin effects in black hole binaries.\cite{Da.01,Da.al2.08,BaBu.10,BaBu.11,Na.11,BaJe.13} The EOB model further incorporates a description of the gravitational-wave emission and the related dissipative radiation-reaction force.\cite{DaNa.08,Da.al.09,BiDa.12} Both conservative and dissipative sectors rely heavily on resummation methods such as Pad\'e approximants,\cite{Da.al.97,Da.al3.01} aimed at improving the convergence of the PN series in the strong-field regime (see however Refs.~\refcite{Bl.02,Bo.al.08,Mr.al.08} and \refcite{Fa.11}). To account for uncontrolled relativistic corrections during the late inspiral and final plunge, the EOB model also makes use of several free parameters that are fitted by comparison to the results of fully nonlinear NR simulations. Ongoing work focuses on calibrating several versions of the model to NR simulations for increasingly generic binary configuations.\cite{Pa.al.11,Ta.al.12,Pa.al.14,Pa.al2.14,Ta.al.14,Da.al.13,DaNa.14} See Refs.~\refcite{DaNa.11,Da.14} and \refcite{Da2.14} for recent reviews.

\vspace{0.25cm}

\begin{figure}
	\begin{center}
		\includegraphics[width=0.65\linewidth]{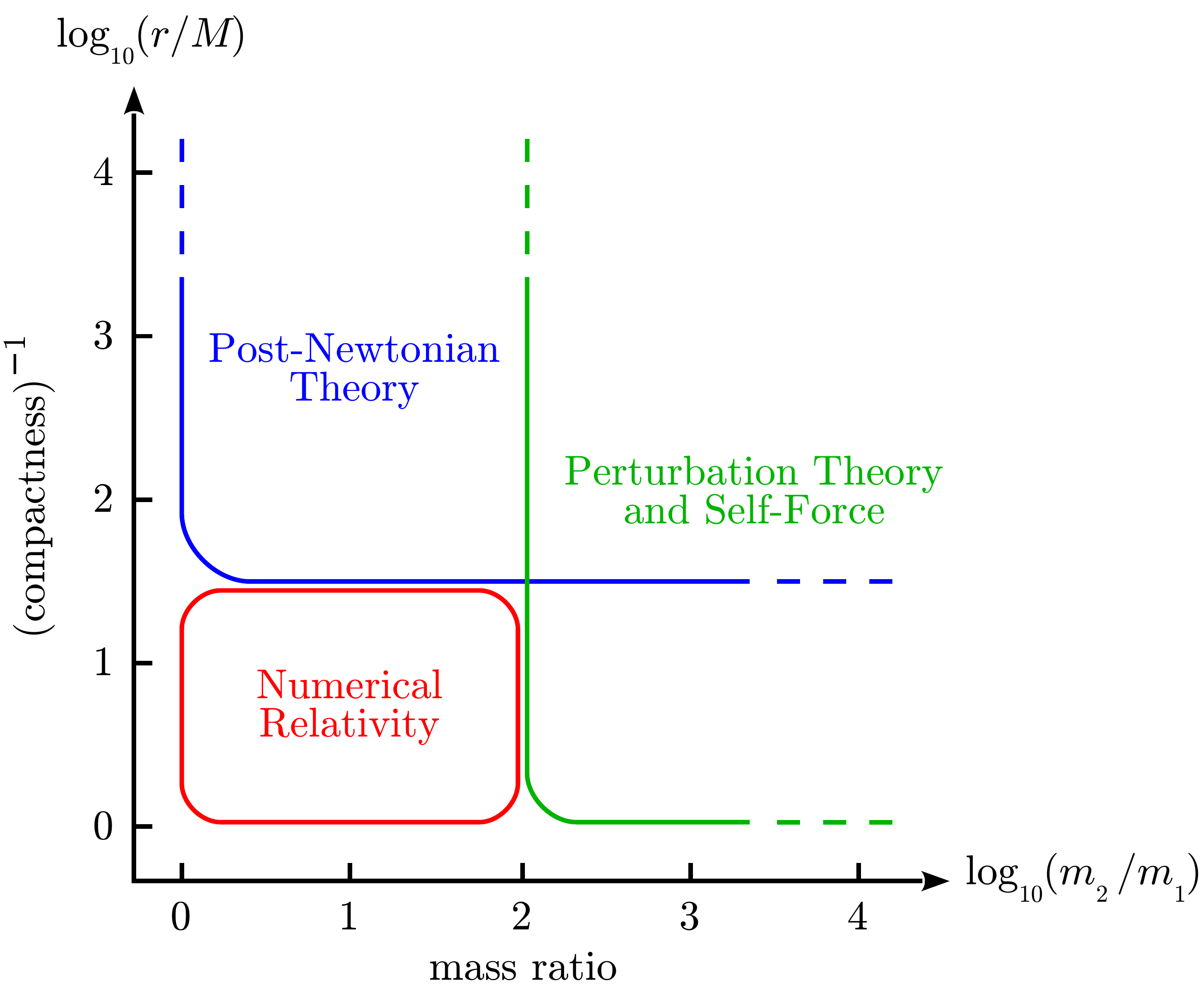}
		\caption{Different analytical approximation schemes and numerical techniques are used to model the orbital dynamics and gravitational-wave emission from black hole binaries, according to the mass ratio $0 < m_1 / m_2 \leqslant 1$ and the compactness parameter $0 < M / r \lesssim 1$, where $M = m_1 + m_2$ is the total mass and $r$ the typical binary separation.}
		\label{fig:methods}
	\end{center}
\end{figure}

These approximation methods and numerical techniques are depicted in Fig.~\ref{fig:methods}. While the domain of validity of NR simulations does, in principle, cover the entire parameter space, in practice it is constrained by available computational ressources. Indeed, both wide separations and large mass ratios require exceedingly long computations. The domains of validity of PN theory and BHP theory are not delimited by sharp boundaries either; these depend on the acceptable level of error made in approximating the exact result for any given calculation. Borrowing results from the PN approximation and BHP theory, as well as nonperturbative information from NR simulations, the EOB model aims at covering the entire parameter space.

\vspace{-0.25cm}

\subsection{Black hole perturbations, post-Newtonian expansions and numerical simulations}

It is crucial to compare the predictions from this variety of approximation methods and numerical techniques for several reasons. Indeed, such comparisons (i) provide independent consistency checks of the validity of the various calculations, (ii) they help to delineate the respective domains of validity of each approximation method, and (iii) they inform the development of a universal, semi-analytical model of the binary dynamics and gravitational-wave emission.

Since Einstein's equation is covariant under general coordinate transformations, it is best to perform such comparisons by using \textit{coordinate-invariant} relationships, whose functional form is independent of the particular choice of coordinates used to perform the calculation. This circumvents the need to ensure that the same coordinate system is being used in all calculations. Prime examples of such relationships are the gravitational-wave polarizations $h_+(t)$ and $h_\times(t)$ as functions of the proper time $t$ of an asymptotically far inertial observer, as well as physical quantities derived from them, such as the gravitational-wave fluxes of energy and linear momentum. Several comparisons relying on the radiative aspects of binary black hole spacetimes will be discussed in Secs.~\ref{sec:h} and \ref{sec:F}.

Moreover, in the context of approximation schemes such as PN theory and BHP theory, one can perform a meaningful split between the conservative and dissipative parts of the orbital dynamics, the latter being associated with the emission of gravitational radiation.\footnote{Such as split can be meaningfully defined up to \textit{at least} fourth order in the PN approximation\cite{Da.al.14} and second order in BHP theory.\cite{Po.14}} Sections \ref{sec:z} and \ref{sec:psi} will review two recent comparisons of these approximation methods, based on \textit{conservative} effects on the dynamics of black hole binaries moving along circular orbits. In general, such a split cannot be achieved in full general relativity. However, as long as dissipative effects remain sufficiently small, (approximately) conservative effects on the dynamics of black hole binaries can be extracted from NR simulations. The comparisons reviewed in Secs.~\ref{sec:K} and \ref{sec:E}, which probe different aspects of the conservative dynamics, will include NR results for two black holes moving along an \textit{adiabatic} sequence of quasi-circular orbits.

Tables \ref{tab:waveform} and \ref{tab:dynamics} collect recent work (since 2007) featuring such comparisons.\footnote{References discussing NR/EOB comparisons of the predicted waveforms that involve a \textit{calibration} of the EOB model to the results of NR simulations are not listed in Table \ref{tab:waveform}.} The observables used to compare the predictions of the various methods are listed in each case. Most of these comparisons are restricted to the simplest case of non-spinning binary black holes moving on quasi-circular orbits, but some authors have considered spin effects or different types of orbits (\textit{e.g.} quasi-eccentric and head-on). Thereafter, we shall review only a sample of this large body of work, emphasizing comparisons relying on NR simulations, PN expansions and black hole perturbations; numerous comparisons involving the predictions of the EOB model are discussed extensively in the reviews \refcite{DaNa.11,Da.14} and \refcite{Da2.14}.

\begin{table}[h!]
	\renewcommand{\arraystretch}{1.1}
	\tbl{Recent comparisons of the predictions from numerical relativity simulations (NR), black hole perturbation theory (BHP), the post-Newtonian approximation (PN), and the effective one-body model (EOB), based on the computed gravitational waveforms or derived quantities.}
	{\begin{tabular}{@{}lllllc@{}}
		\toprule
		\textbf{Paper} & \textbf{Year} & \textbf{Methods} & \textbf{Observable} & \textbf{Orbit} & \textbf{Spin} \\
		\colrule
		Baker \textit{et al.}\cite{Ba.al.07} & 2007 & NR/PN & waveform & & \\
		Boyle \textit{et al.}\cite{Bo.al2.07} & 2007 & NR/PN & waveform & & \\
		Hannam \textit{et al.}\cite{Ha.al.07} & 2007 & NR/PN & waveform & & \\
		Boyle \textit{et al.}\cite{Bo.al.08} & 2008 & NR/PN/EOB & energy flux & & \\
		Damour \& Nagar\cite{DaNa.08} & 2008 & NR/EOB & waveform & & \\
		Hannam \textit{et al.}\cite{Ha.al2.08} & 2008 & NR/PN & waveform & & \ding{51} \\
		Pan \textit{et al.}\cite{Pa.al.08} & 2008 & NR/PN/EOB & waveform & & \\
		Campanelli \textit{et al.}\cite{Ca.al.09} & 2009 & NR/PN & waveform & & \ding{51} \\
		Hannam \textit{et al.}\cite{Ha.al.10} & 2010 & NR/PN & waveform & & \ding{51} \\
		Hinder \textit{et al.}\cite{Hi.al.10} & 2010 & NR/PN & waveform & eccentric & \\
		Lousto \textit{et al.}\cite{Lo.al2.10} & 2010 & NR/BHP & waveform & & \\
		Sperhake \textit{et al.}\cite{Sp.al.11} & 2011 & NR/PN & waveform & & \\
		Sperhake \textit{et al.}\cite{Sp.al2.11} & 2011 & NR/BHP & waveform & head-on & \\
		Lousto \& Zlochower\cite{LoZl.11} & 2011 & NR/BHP & waveform & & \\
		Nakano \textit{et al.}\cite{Na.al.11} & 2011 & NR/BHP & waveform & & \\
		Lousto \& Zlochower\cite{LoZl.13} & 2013 & NR/PN & waveform & & \\
		Nagar\cite{Na.13} & 2013 & NR/BHP & recoil velocity & & \\
		Hinder \textit{et al.}\cite{Hi.al.14} & 2014 & NR/PN/EOB & waveform & & \ding{51} \\
		\botrule
	\end{tabular}\label{tab:waveform}}
\end{table}
\begin{table}[h!]
	\renewcommand{\arraystretch}{1.1}
	\tbl{Same as in Table \ref{tab:waveform}, but using coordinate-invariant diagnostics of the conservative part of the circular-orbit dynamics instead of the gravitational waveforms.}
	{\begin{tabular}{@{}llllc@{}}
		\toprule
		\textbf{Paper} & \textbf{Year} & \textbf{Methods} & \textbf{Observable} & \textbf{Spin} \\
		\colrule
		Detweiler\cite{De.08} & 2008 & BHP/PN & redshift observable & \\
		Blanchet \textit{et al.}\cite{Bl.al.10,Bl.al2.10} & 2010 & BHP/PN & redshift observable & \\
		Damour\cite{Da.10} & 2010 & BHP/EOB & ISCO frequency & \\
		Mrou{\'e} \textit{et al.}\cite{Mr.al.10} & 2010 & NR/PN & periastron advance & \\
		Barack \textit{et al.}\cite{Ba.al.10} & 2010 & BHP/EOB & periastron advance & \\
		Favata\cite{Fa.11} & 2011 & BHP/PN/EOB & ISCO frequency & \\
		Le Tiec \textit{et al.}\cite{Le.al.11} & 2011 & NR/BHP/PN/EOB & periastron advance & \\
		Damour \textit{et al.}\cite{Da.al.12} & 2012 & NR/EOB & binding energy & \\
		Le Tiec \textit{et al.}\cite{Le.al2.12} & 2012 & NR/BHP/PN/EOB & binding energy & \\
		Akcay \textit{et al.}\cite{Ak.al.12} & 2012 & BHP/EOB & redshift observable & \\
		Hinderer \textit{et al.}\cite{Hi.al.13} & 2013 & NR/EOB & periastron advance & \ding{51} \\
		Le Tiec \textit{et al.}\cite{Le.al.13} & 2013 & NR/BHP/PN & periastron advance & \hspace{0.01cm} \ding{51} \vspace{0.04cm} \\
		\parbox{2.9cm}{Bini \& Damour\cite{BiDa.13,BiDa.14,BiDa2.14} \\ Shah \textit{et al.}\cite{Sh.al.14} \\ Blanchet \textit{et al.}\cite{Bl.al.14,Bl.al2.14}} $\Biggr\}$ & 2014 & BHP/PN & redshift observable & \vspace{0.15cm} \\
		\parbox{2.4cm}{Dolan \textit{et al.}\cite{Do.al.14} \\ Bini \& Damour\cite{BiDa3.14}} $\Bigr\}$ & 2014 & BHP/PN & precession angle & \hspace{0.01cm} \ding{51} \vspace{0.1cm} \\
		Isoyama \textit{et al.}\cite{Is.al.14} & 2014 & BHP/PN/EOB & ISCO frequency & \ding{51} \\
		\botrule
	\end{tabular}\label{tab:dynamics}}
\end{table}

\section{Gravitational Waveforms}\label{sec:h}

The primary objective of analytical and numerical relativists working on the general relativistic two-body problem is to accurately model the gravitational-wave emission from inspiralling and coalescing compact-object binaries. The polarizations $h_+$ and $h_\times$ are thus often used to compare the predictions from NR simulations, PN theory, BHP theory and the EOB model. To dispense with their angular dependance, it is convenient to perform a mode decomposition of the linear combination $h \equiv h_+ - \ui h_\times$. Far away from the isolated source, and using an asymptotically radiative, or Bondi-type, coordinate system $\{t,r_*,\theta,\phi\}$, we have
\beq\label{h}
	h = \frac{M}{r_*} \sum_{\ell=2}^\infty \sum_{m=-\ell}^\ell h_{\ell m}(u) \, Y^{\ell m}_{-2}(\theta,\phi) \, ,
\eeq
where $u \equiv t - r_*$ is the retarded time, $Y^{\ell m}_{-2}(\theta,\phi)$ are spin-weighted spherical harmonics of spin $s = -2$, and terms $\calO(r_*^{-2})$ are neglected. Alternatively, the output of NR simulations is often given in terms of the Weyl scalar $\Psi_4$ which, far away from the source, is related to the wave polarizations through\footnote{Several conventions for the definition of the Weyl scalar are commonly used in the NR litterature, such that the alternative relationships $\Psi_4 = \pm \frac{1}{2} \ddot{h}$ can often be found.} $\Psi_4 = \partial^2_t h \equiv \ddot{h}$. Therefore, one is often led to compute the modes $\psi_{\ell m}$ of $\Psi_4$, which are simply related to the modes $h_{\ell m}$ of $h$ by $\psi_{\ell m} = M^2 \ddot{h}_{\ell m}$.

In the astrophysically relevant case of the quasi-circular inspiral of two compact objects, the modes $h_{\ell m}$ have been computed up to high orders in the PN approximation.\cite{Bl.al.96,Ar.al.04,Ki.al.07,Ki.08,Bl.al.08} For instance, for non-spinning bodies, the (dominant) quadrupolar mode $h_{22}$ is known including all amplitude corrections up to 3.5PN order,\cite{Fa.al.12} and reads\footnote{We use the standard redefinition of the orbital phase variable $\Phi$ to absorb some logarithmic terms resulting from tail effects.\cite{Bl.al.96,Fa.al.12}}
\beq\label{h_22}
	h_{22} = \sqrt{\frac{64\pi}{5}} \, \nu \, x \, \biggl\{ 1 + \sum_{n=2}^7 \sum_{k=0}^1 a_{n,k}(\nu) \, x^{n/2} \, (\ln{x})^k + o(x^{7/2}) \biggr\} \, e^{-2\ui \Phi} \, ,
\eeq
where $\Phi(t)$ is the orbital phase and $x \equiv (M \Omega)^{2/3}$ the usual dimensionless, invariant PN parameter $\calO(c^{-2})$, with $\Omega(t) \equiv \ud \Phi / \ud t$ the orbital frequency. The coefficients $a_{n,k}(\nu)$ are polynomials in the \textit{symmetric mass ratio} $\nu \equiv \mu / M = m_1 m_2 / (m_1 + m_2)^2$, such that $\nu = 1/4$ for equal-mass binaries and $\nu \to 0$ in the extreme mass-ratio limit. Their expressions can be found in Eq. (6.5) of Ref.~\refcite{Fa.al.12}. The logarithmic running in Eq.~\eqref{h_22} appears at leading 3PN order, \textit{i.e.}, $a_{n,1} = 0$ for $n < 6$.

The modes $h_{\ell m}(t)$ depend on time through the orbital phase $\Phi(t)$ and frequency $\Omega(t)$. If the typical radiation-reaction timescale $T_\text{r.r.}$ is much longer than the typical orbital timescale $T_\text{orb}$, then $\dot{\Omega} / \Omega^2 \sim T_\text{orb} / T_\text{r.r.} \ll 1$ and the slow inspiralling motion can be approximated by an adiabatic sequence of circular orbits. Since the dissipative effects of radiation reaction appear at leading 2.5PN order, $\dot{\Omega} / \Omega^2 = \mathcal{O}(x^{5/2})$, this approximation remains valid until the PN expansion itself breaks down ($x \lesssim 1$). The phasing of the waves can then be computed by means of an argument of energy balance, which states that the mechanical energy of the binary (see Sec.~\ref{sec:E}) decreases at a rate precisely given by the gravitational-wave flux of energy (see Sec.~\ref{sec:F}). There are several ways to compute the phasing $\Phi(t)$ from the equation of energy balance, yielding different so-called \textit{Taylor approximants}.\cite{Da.al3.01,Bo.al2.07} These formally agree up to a given PN order, but differ in how higher-order (uncontrolled) terms are truncated.

\begin{figure}
	\begin{center}
		\includegraphics[width=\linewidth]{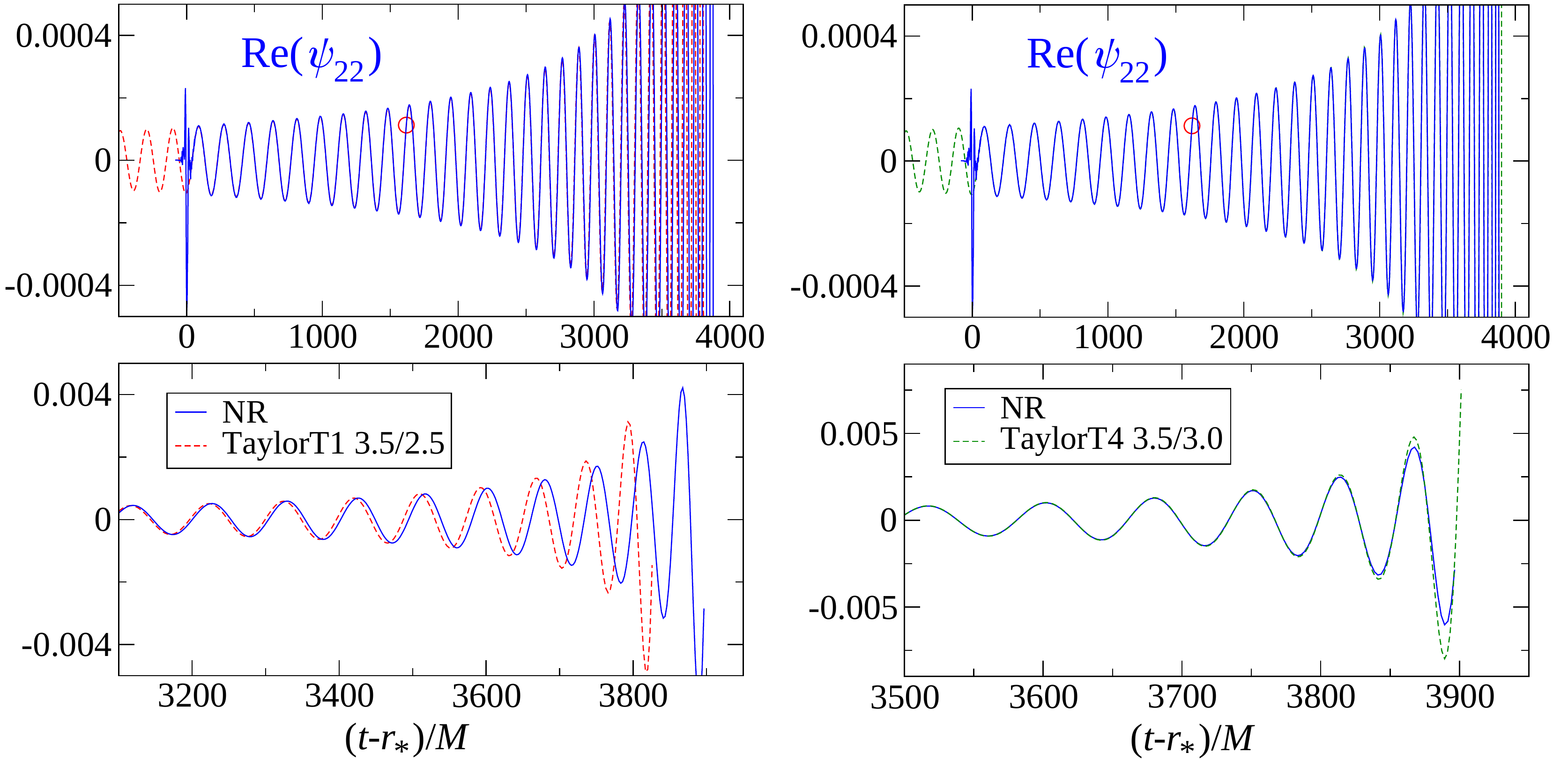}
		\caption{The mode $\psi_{22}$ of the Weyl scalar $\Psi_4$ for the late quasi-circular inspiral of an equal-mass, non-spinning binary black hole, as computed using NR simulations (solid blue) and two PN Taylor approximants (dashed red and dahsed green). \textit{Reproduced from Ref.~\protect\refcite{Bo.al2.07}.}}
		\label{fig:psi_q1}
	\end{center}
\end{figure}

Many studies have investigated the performance of existing Taylor approximants by comparison to NR simulations of non-spinning binaries,\cite{Ba.al.07,Bo.al2.07,Ha.al.07,DaNa.08,Pa.al.08,Sp.al.11,LoZl.13} as well as non-precessing\cite{Ha.al2.08,Ha.al.10} and precessing\cite{Ca.al.09,Hi.al.14} spinning systems. For example, Fig.~\ref{fig:psi_q1} shows the (dominant) mode $\psi_{22}$ of the Weyl scalar $\Psi_4$ in the simplest case of the late quasi-circular inspiral of an equal-mass, non-spinning binary system, as computed using NR simulations (solid blue) and two Taylor approximants (dashed red and dashed green): TaylorT1 with 2.5PN amplitude corrections and TaylorT4 with 3PN amplitude corrections, both using a 3.5PN-accurate phase evolution. For this system, matching numerical results to PN waveforms early in the inspiral yields excellent agreement over the first $\sim 15$ gravitational-wave cycles, thus validating the numerical simulations and establishing a regime where PN theory is accurate. In the last $\sim 15$ cycles to merger, however, Taylor approximants typically build up phase differences of several radians.\footnote{The excellent performance of the TaylorT4 approximant in the case of non-spinning, equal-mass binaries, as clearly visible in Fig.~\ref{fig:psi_q1}, is most likely accidental.\cite{Ha.al2.08,Sp.al.11}} These findings hold true for more generic binary configurations (unequal masses and nonzero spins).

\begin{figure}
	\begin{center}
		\includegraphics[width=\linewidth]{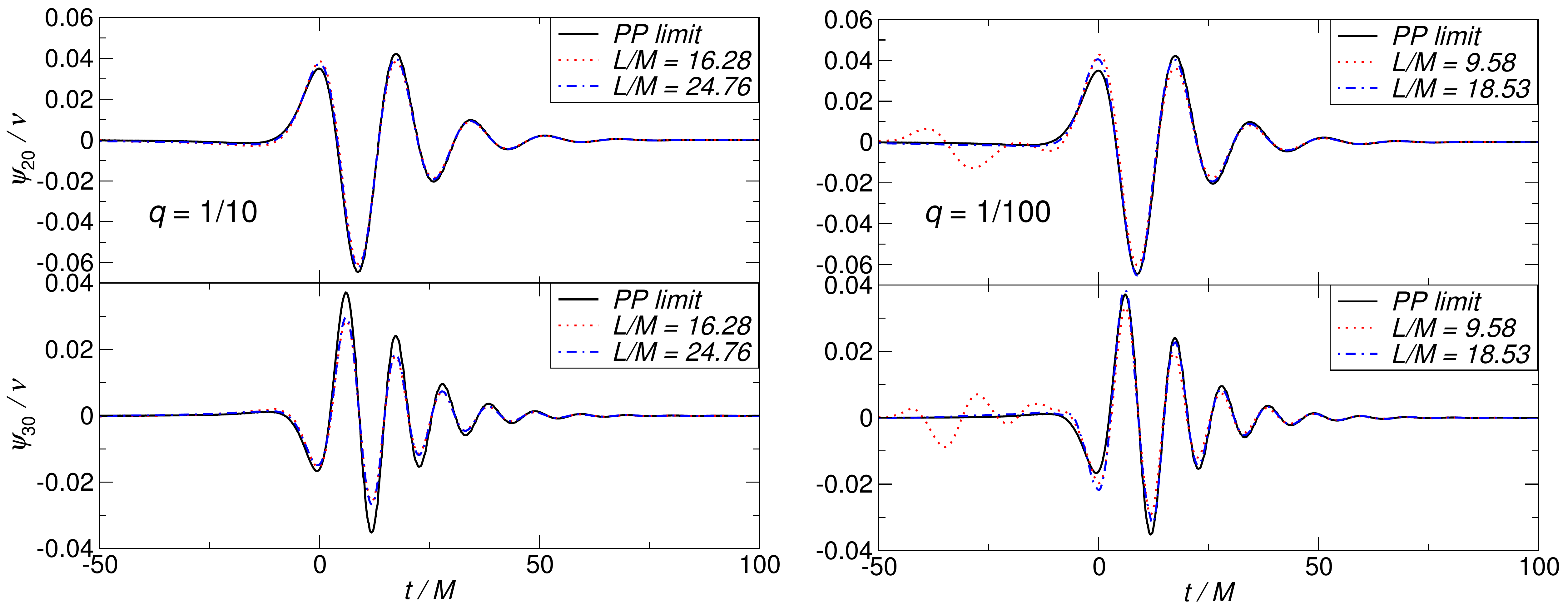}
		\caption{The modes $\psi_{20}$ and $\psi_{30}$ of the Weyl scalar $\Psi_4$, rescaled by the symmetric mass ratio $\nu$, for the head-on collision of two non-spinning black holes with mass ratios $q = 1/10$ (left panel) and $1/100$ (right panel), as computed using NR simulations (dotted red and dashed-dotted blue) and BHP theory to leading order (solid black). \textit{Reproduced from Ref.~\protect\refcite{Sp.al2.11}.}}
		\label{fig:psi_q10}
	\end{center}
\end{figure}

On the other hand, Ref.~\refcite{Sp.al2.11} revisited the classical problem of the head-on collision of two black holes in the small mass-ratio regime.\cite{Da.al.71} Using NR simulations, the authors computed the gravitational waveform generated by two spinless black holes with mass ratios $q = 1/4$, $1/10$ and $1/100$, for several initial proper separations $L$, and compared the results to the standard prediction from linear BHP theory for a test mass, initially at rest, falling radially from infinity into a Schwarzschild black hole.\cite{Da.al.71} Figure \ref{fig:psi_q10} shows the modes $\psi_{20}$ and $\psi_{30}$ of the Weyl scalar $\Psi_4$, rescaled by the symmetric mass ratio $\nu = q / (1+q)^2 = q + \calO(q^2)$, for mass ratios $q = 1/10$ and $1/100$. Note the remarkable agreement between the NR results and the leading-order rescaled ($q \to \nu$) perturbative prediction, despite the intermediate mass ratios. Earlier work comparing the predictions of NR and BHP theory already suggested the weak dependence of the rescaled waveform $\psi_{\ell m} / \nu$ on the mass ratio $q$.\cite{De.79,Sm.79,An.al.95}. The findings of Ref.~\refcite{Sp.al2.11} confirm this observation over a wide range of mass ratios.

\vspace{0.2cm}

\section{Fluxes of Energy and Linear Momentum}\label{sec:F}

As previously mentioned, one of the key ingredients in the construction of templates for compact-object binaries is the gravitational-wave flux of energy, or luminosity, $\calF$. From the effective stress-energy tensor asociated with gravitational radiation,\cite{Is1.68,Is2.68} the energy flux can be computed from the knowledge of the waveform, by performing a surface integral over a two-sphere at future null infinity:
\beq
	\calF = \frac{1}{16\pi} \lim_{\stackrel{r_* \to \infty}{u = \text{cst}}} \oint \vert \dot{h} \vert^2 \, \ud A \, ,
\eeq
where $\ud A = r_*^2 \, \ud \Omega$. For two non-spinning compact objects moving along an adiabatic sequence of quasi-circular orbits with time-varying frequency $\Omega(t)$, the coordinate-invariant relation $\calF(\Omega)$ has been computed at the relative 3.5PN accuracy.\cite{Bl.al.02,Bl.al2.02,Bl.al2.04,Bl.al2.05} The result, which is valid for arbitrary mass ratios, reads [recall that $x = (M \Omega)^{2/3}$]
\beq\label{F_PN}
	\calF = \frac{32}{5} \, \nu^2 x^5 \, \biggl\{ 1 + \sum_{n=2}^7 \sum_{k=0}^1 b_{n,k}(\nu) \, x^{n/2} \, (\ln{x})^k + o(x^{7/2}) \biggr\} \, ,
\eeq
where the coefficients $b_{n,k}(\nu)$ are polynomials in the symmetric mass ratio $\nu$. Their expressions can be found, \textit{e.g.}, in Eq. (173) of Ref.~\refcite{Bl.14}. The leading-order term in \eqref{F_PN} originates from the application of Einstein's quadrupole formula to a binary system of point masses moving on a Keplerian circular orbit.

On the other hand, for a point particle of mass $m_1$ orbiting a Schwarzschild hole of mass $m_2 \gg m_1$, the Regge-Wheeler and Teukolsky master equations\cite{ReWh.57,Te.72} for the metric perturbation and Weyl scalar $\Psi_4$ can be solved analytically using the perturbative techniques developped in Refs.~\refcite{Ma.al.96} and \refcite{MaTa.97}. A large body of work\cite{Ta.al.93,TaSa.94,Ta.al2.96,Mi.al2.97} by the Japanese relativity school has culminated in the calculation of the luminosity $\calF$, for quasi-circular orbits, at the outstanding 22PN relative accuracy,\cite{Fu.12,Fu2.12} and could in principle be extended to arbitrarily high PN orders.\footnote{The perturbative gravitational-wave flux of energy has also be computed numerically, with high accuracy, without any weak-field expansion, yielding fully relativistic results.\cite{Cu.al2.93,TaNa.94,Sh.94,Hu.00,FuTa.04}} The result is valid at leading order in the mass ratio $q = m_1 / m_2$, and reads
\beq\label{F_BH}
	\calF = \frac{32}{5} \, q^2 y^5 \, \biggl\{ 1 + \sum_{n \geqslant 2} \sum_{k \geqslant 0} c_{n,k} \, y^{n/2} \, (\ln{y})^k + \calO(q) \biggr\} \, ,
\eeq
where $y \equiv (m_2 \Omega)^{2/3}$ is a frequency-related parameter $\calO(c^{-2})$ that appears naturally in the context of BHP theory. The numerical coefficients $c_{n,k}$ can be found, \textit{e.g.} up to 5.5PN order, in Eq.~(174) of Ref.~\refcite{SaTa.03}. The extreme mass-ratio limit of the 3.5PN prediction \eqref{F_PN} and the 3.5PN restriction of the perturbative result \eqref{F_BH} are in perfect agreement, \textit{i.e.}, $b_{n,k}(0) = c_{n,k}$ for $2 \leqslant n \leqslant 7$ and $0 \leqslant k \leqslant 1$. Note that the correct, leading-order scaling in the masses can easily be reproduced from the perturbative result by setting $q \to \nu = q / (1+q)^2$ and $y \to x = y \, (1+q)^{2/3}$ in Eq.~\eqref{F_BH}.

This agreement extends to spinning bodies as well, for which the spin-orbit contributions to the energy flux (\textit{i.e.} the terms linear in the spins) have been computed both in the PN approximation, up to the next-to-next-to-leading order\cite{Ki.95,Bl.al.06,Bo.al.13} (corresponding to 3.5PN order for large spins), and in BHP theory for a particle on a circular equatorial orbit around a Kerr black hole.\cite{Po2.93,Sh.al.95,Ta.al.96,Sh.14}

Gravitational waves do not only carry energy away from their source; they also carry linear momentum. Therefore, by conservation of the total linear momentum, the Kerr black hole resulting from the coalescence of two compact bodies must have a recoil (or kick) velocity $\bm{v}$ with respect to the center-of-mass frame, given by
\beq
	\bm{v} = - \frac{1}{M} \int \bm{\calF}(t) \, \ud t \, ,
\eeq
where the gravitational-wave flux of linear momentum may also be computed from the knowledge of the modes $h_{\ell m}$ of the waveform: $\bm{\calF} = \lim_{r_* \to \infty} \oint \vert \dot{h} \vert^2 \, \bm{n} \, \frac{\ud A}{16\pi}$, with $\bm{n}$ the unit radial vector in flat space.\cite{Ru.al.08} This \textit{gravitational recoil} effect has received a lot of attention due to its potential astrophysical consequences.\cite{Me.al.04} For non-spinning binary black holes, a number of analytical estimate of the final recoil velocity $v \equiv |\bm{v}|$ have been given both prior\cite{Fi.83,FiDe.84,Bl.al.05,So.al.06,DaGo.06,LeBl.10,Le.al.10} and after\cite{Re.al.10,Mi.al.12,Mi.12} its extensive exploration by means of NR simulations.\cite{Ba.al3.06,Go.al.07,He.al.07,Ko.al.07,Go.al.09,Bu.al.12,Ja.al.12}

\begin{figure}
	\begin{center}
		\includegraphics[width=0.8\linewidth]{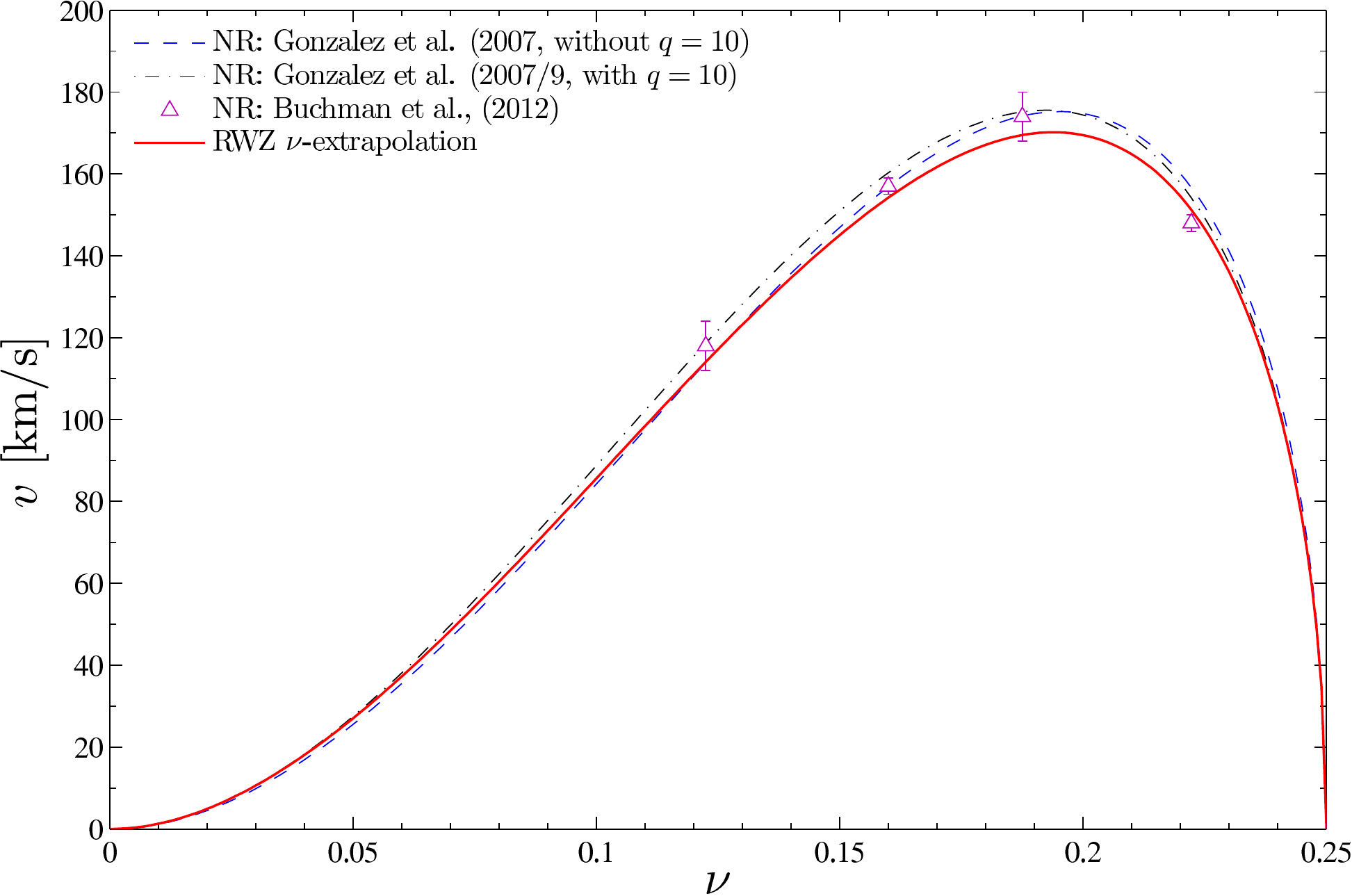}
		\caption{The recoil velocity $v = |\bm{v}|$ as a function of the symmetric mass ratio $\nu$, as computed for non-spinning black hole binaries using NR simulations (dashed blue, dashed-dotted black, magenta triangles) and linear BHP theory with a $\nu$-rescaling of the modes of the gravitational waveform. \textit{Reproduced from Ref.~\protect\refcite{Na.13}}.}
		\label{fig:v}
	\end{center}
\end{figure}

In particular, Nagar\cite{Na.13} showed that extrapolating in $\nu$ the waveform obtained in the limit $q \ll 1$ via a perturbative approach, multipole by multipole (up to multipole order $\ell = 8$), and then computing the recoil velocity $v$ from this $\nu$-flexed waveform, yields a prediction that is compatible with the results from NR simulations. Figure \ref{fig:v} shows the remarkable agreement between this $\nu$-rescaled perturbative calculation (solid red) and the exact results (magenta triangles), as computed using NR simulations of \textit{comparable-mass} binary black holes.\cite{Bu.al.12} The dashed blue and dashed-dotted black curves show fits to existing NR results.\cite{Go.al.07,Go.al.09}

\vspace{0.2cm}

\newpage
	
\section{Redshift Observable}\label{sec:z}

Next, we turn to a comparison of the predictions from the PN approximation and GSF theory that relies on a \textit{conservative} effect on the orbital dynamics of a binary system of non-spinning compact bodies moving on a circular orbit. For such systems, the metric $g_{ab}$ is invariant along the integral curves of a \textit{helical} Killing field $k^a$, which can be normalized such that $k^a \to {(\partial_t)}^a + \Omega \, {(\partial_\varphi)}^a$ at infinity, where the constant $\Omega$ has the interpretation of the angular velocity of the orbit. 

Within both approximation schemes, at least one of the (non-spinning) compact objects can be modelled as a point mass obeying \textit{geodesic} motion in some suitably regularized metric.\cite{Bl.14,Sc.11,Po.al.11} Let $\gamma$ denote the timelike worldline of the lightest body, with unit tangent $u^a$. Then, at the particle the helical Killing vector is tangent to the four-velocity: $k^a|_\gamma = z \, u^a$. Using the normalization condition $u^a u_a = -1$, the coefficient of proportionality is given by
\beq\label{z}
	z^2 = - k^a k_a \vert_\gamma \, ,
\eeq
or alternatively $z = - k^a u_a$. Since the worldline $\gamma$ is a geodesic, $z$ is the Killing energy of the particle, a conserved quantity associated with the helical symmetry. Because the scalar $z$ can also be interpreted as the gravitational redshift of light emitted from the particle and received far away from the binary, along the helical symmetry axis,\cite{De.08} it is often referred to as the ``redshift observable.''\footnote{Despite its name, the redshift observable does not account for the gravitational redshift induced by the self-field of the body. Indeed, while computing $z$, the singular self-field of the point mass is subtracted from the physical metric in both the PN and GSF formalisms.\cite{Bl.al.10}}  Using coordinates adapted to the symmetry, \textit{i.e.}, such that $k^a = {(\partial_t)}^a + \Omega \, {(\partial_\varphi)}^a$ everywhere, the redshift observable reads simply:
\beq
	z = \frac{1}{u^t} = \frac{\ud \tau}{\ud t} \, .
\eeq
It coincides with the ratio of proper times elapsed along the worldlines of the particle and of an asymptotically far static observer. The relationship $z(\Omega)$ being coordinate-invariant, it can be used to perform a meaningful comparison of the predictions from the perturbative GSF formalism and the PN approximation.

In the context of PN theory, the relation $z(\Omega)$ was computed up to 2PN and then 3PN order in Refs.~\refcite{De.08} and \refcite{Bl.al.10}. The leading order and next-to-leading order logarithmic contributions at 4PN and 5PN orders, which originate from gravitational-wave tails, were obtained in Refs.~\refcite{Bl.al2.10,Da.10} and \refcite{Le.al.12}. The non-logarithmic 4PN terms are also known from the 4PN contribution to the circular-orbit binding energy\cite{JaSc.13,BiDa.13} (cf. Sec.~\ref{sec:E} below) and application of the first law of binary black hole mechanics.\cite{Le.al.12} All of these results are valid for any mass ratio.\footnote{The leading, next-to-leading, and next-to-next-to-leading half-integral terms at 5.5PN, 6.5PN and 7.5PN orders, which originate from gravitational-wave tails of tails, were also computed, but only at linear order in the mass ratio.\cite{Bl.al.14,Bl.al2.14}} Collecting them up to 4PN order, the redshift of the lightest body reads
\begin{align}\label{z_PN}
	z &= 1 + \left( - \frac{3}{4} - \frac{3}{4} \Delta + \frac{\nu}{2} \right) x + \left( - \frac{9}{16} - \frac{9}{16} \Delta - \frac{\nu}{2} - \frac{1}{8} \Delta \, \nu + \frac{5}{24} \nu^2 \right) x^2 \nonumber \\ &\qquad\! + \left( - \frac{27}{32} - \frac{27}{32} \Delta - \frac{\nu}{2} + \frac{19}{16} \Delta \, \nu - \frac{39}{32} \nu^2 - \frac{1}{32} \Delta \, \nu^2 + \frac{\nu^3}{16} \right) x^3 \nonumber \\ &\qquad\! + \left( - \, \frac{405}{256} - \frac{405}{256} \Delta + \left[ \frac{38}{3} - \frac{41}{64} \pi^2 \right] \nu + \left[ \frac{6889}{384} - \frac{41}{64} \pi^2 \right] \Delta \, \nu \right. \nonumber \\ &\qquad\qquad\!\! \left. + \left[ - \frac{3863}{576} + \frac{41}{192} \pi^2 \right] \nu^2 - \frac{93}{128} \Delta \, \nu^2 + \frac{973}{864} \nu^3 - \frac{7}{1728} \Delta \, \nu^3 + \frac{91}{10368} \nu^4 \right) x^4 \nonumber \\ &\qquad\! + \left( - \, \frac{1701}{512} - \, \frac{1701}{512} \Delta + \left[ - \frac{329}{15} + \frac{1291}{1024} \pi^2 + \frac{64}{5} \gamma_\text{E} + \frac{32}{5} \ln{(16x)} \right] \nu  \right. \nonumber \\ &\qquad\qquad\!\! + \left[ - \frac{24689}{3840} + \frac{1291}{1024} \pi^2 + \frac{64}{5} \gamma_\text{E} + \frac{32}{5} \ln{(16x)} \right] \Delta \, \nu + \left[ - \frac{71207}{1536} + \frac{451}{256} \pi^2 \right] \Delta \, \nu^2 \nonumber \\ &\qquad\qquad\!\! + \left[ - \frac{1019179}{23040} + \frac{6703}{3072} \pi^2 + \frac{64}{15} \gamma_\text{E} + \frac{32}{15} \ln{(16x)} \right] \nu^2 + \left[ \frac{356551}{6912} - \frac{2255}{1152} \pi^2 \right] \nu^3 \nonumber \\ &\qquad\qquad\!\! \left. + \,\, \frac{43}{576} \, \Delta \, \nu^3 - \frac{5621}{41472} \, \nu^4 + \frac{55}{41472} \, \Delta \, \nu^4 - \frac{187}{62208} \, \nu^5 \right) x^5 + o(x^5) \, ,
\end{align}
where $\Delta \equiv (m_2 - m_1) / M = \sqrt{1-4\nu}$ is the reduced mass difference (with $m_1 \leqslant m_2$) and $\gamma_\text{E} = 0.577215\cdots$ is the Euler-Mascheroni constant. The redshift of the heaviest body is simply obtained by setting $\Delta \to - \Delta$ in Eq.~\eqref{z_PN}.

In the context of BHP theory, for a particle of mass $m_1$ orbiting a Schwarzschild black hole of mass $m_2 \gg m_1$ on a circular orbit with angular velocity $\Omega$, the redshift $z(\Omega)$ is Taylor-expanded in powers of the small mass ratio $q = m_1 / m_2$ as
\beq\label{z_BH}
	z = z_\text{Schw}(y) + q \, z_\text{GSF}(y) + \calO(q^2) \, ,
\eeq
where $y = (m_2 \Omega)^{2/3} \equiv m_2 / r_\Omega$ is an invariant measure of the (inverse) orbital radius. The result in the test-particle limit $m_1 \to 0$ is known exactly as $z_\text{Schw}(y) = \sqrt{1-3y}$. It vanishes at the Schwarzschild lightring, the innermost circular orbit at $r_\Omega = 3m_2$, since $\ud \tau \to 0$ for this null geodesic. The limit $\nu \to 0$ (and $\Delta \to 1$) of the 4PN formula \eqref{z_PN} is in agreement with the 4PN expansion of this test-mass result.

The conservative gravitational self-force correction $q \, z_\text{GSF}(\Omega)$ to the test-particle result is essentially given by the double contraction of the regularized metric perturbation $h_{ab}^R$ with the particle's four-velocity $u^a$.\cite{De.08} It has been computed numerically, with high accuracy, by several independent groups using a variety  of formulations, gauge conditions, and numerical methods.\cite{De.08,Sa.al.08,Bl.al.10,Bl.al2.10,Sh.al.11,Ak.al.12} In the strong-field regime, it has the asymptotic behavior $z_\text{GSF}(y) \sim {(1-3y)}^{-1}$ when $y \to 1/3$.\cite{Ak.al.12} In the weak-field regime $y \ll 1$, a number of comparisons with the PN prediction --- as derived by substituting $x = y \, (1+q)^{2/3}$ and $\nu = q / (1+q)^2$ in \eqref{z_PN} and expanding in powers of $q$ --- show a very good agreement,\cite{De.08,Bl.al.10,Bl.al2.10} thus providing a crucial test of the regularization schemes used in PN theory and GSF theory; see Ref.~\refcite{Bl.al.11} for a review.

Recently, using perturbative methods valid to linear order in the mass ratio, even higher order PN coefficients entering the weak-field expansion of $z_\text{GSF}(\Omega)$ have been computed. In particular, Refs.~\refcite{BiDa.13,BiDa.14} and \refcite{BiDa2.14} used the improved BHP techniques of Mano, Suzuki and Takasugi\cite{Ma.al.96,MaTa.97} to analytically compute the PN expansion of $z_\text{GSF}(\Omega)$ up to 8.5PN order. Using similar techniques, the authors of reference \refcite{Sh.al.14} performed an extremely accurate numerical calculation (with 255 significant digits!) of $z_\text{GSF}(\Omega)$, which allowed them to extract the exact, analytical values of many PN coefficients up to 10PN order, and to numerically determine the others with no less than 13 significant digits. These results have uncovered the unexpected occurence of conservative terms at \textit{half-integral} PN orders, which correspond to odd powers of $1/c$, beginning at 5.5PN order; these terms are now understood to originate from gravitational-wave tails of tails.\cite{Bl.al.14,Bl.al2.14}

\begin{figure}
	\begin{center}
		\includegraphics[width=0.85\linewidth]{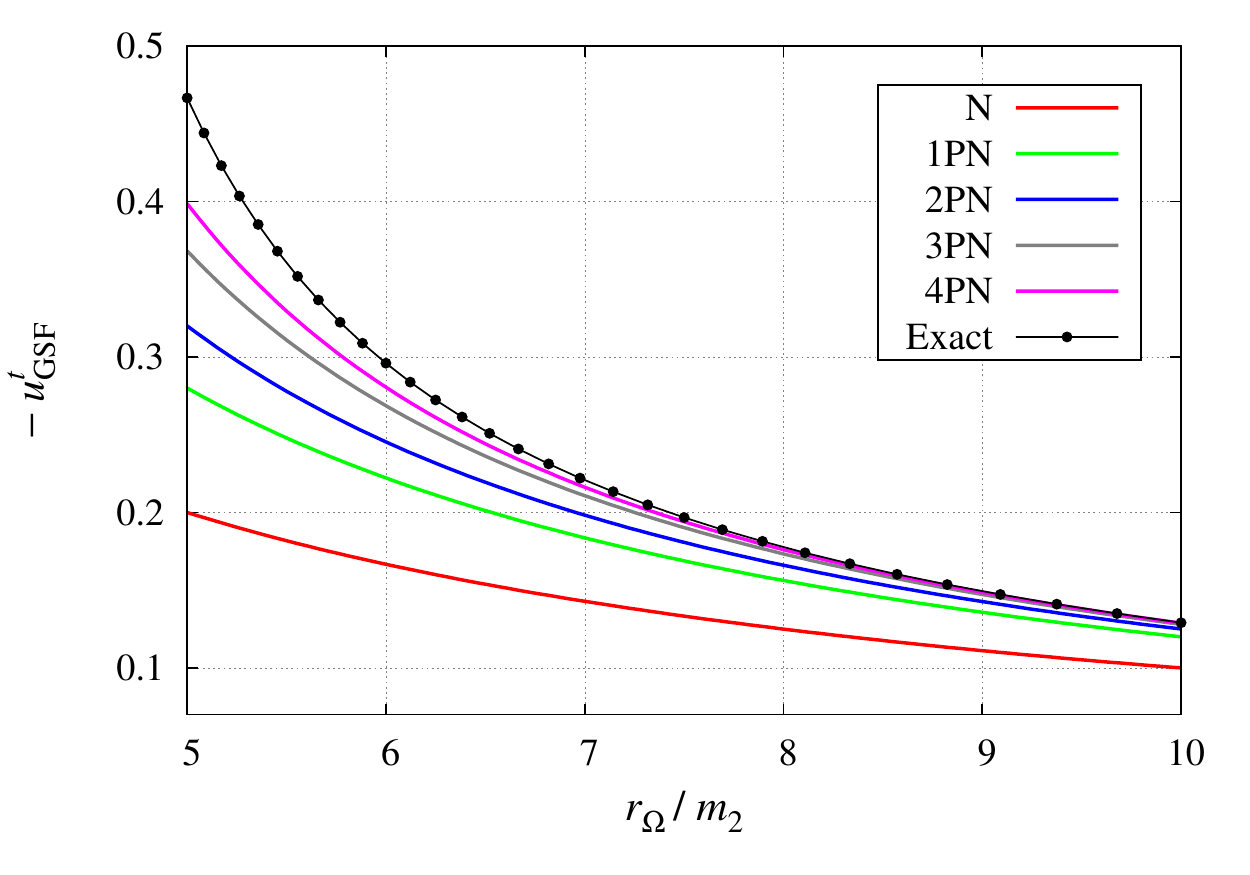}
		\caption{The conservative gravitational self-force contribution $u^t_\text{GSF}$ to the redshift observable as a function of $r_\Omega \equiv (m_2 / \Omega^2)^{1/3}$, a coordinate-invariant measure of the orbit separation, as computed numerically in BHP theory and analytically in PN theory up to 4PN order. Notice that $r_\Omega = 6 m_2$ corresponds to the Schwarzschild innermost stable circular orbit (ISCO). \textit{Reproduced from Ref.~\protect\refcite{Bl.al2.10}.}}
		\label{fig:z}
	\end{center}
\end{figure}

Summarizing the results of Refs.~\refcite{De.08,Bl.al.10,Bl.al2.10,Sh.al.14,BiDa.13,BiDa.14,BiDa2.14,Bl.al.14} and \refcite{Bl.al2.14}, the weak-field expansion of the conservative GSF contribution to $u^t = 1/z$ reads\footnote{The general form of the near-zone PN expansion is known to be of the type $\sum_{n,k} (\ln{c})^k/c^n$.\cite{BlDa.86}}
\beq\label{ut_GSF}
	u^t_\text{GSF}(y) = \sum_{n \geqslant 0} \left( \alpha_n + \beta_n \ln{y} + \gamma_n \ln^2{y} + \cdots \right) y^{n+1} ,
\eeq
where $n$ can take both integral and half-integral values. The analytically determined PN coefficients $\alpha_n$, $\beta_n$ and $\gamma_n$ are summarized in Table~\ref{tab:coeffs}, up to 7.5PN order. Note that the transcendentality of the coefficients $\alpha_n$ increases with $n$. The coefficients in bold font were tested against \textit{independent} PN calculations valid for arbitrary mass ratios.\cite{De.08,Bl.al.10,Bl.al2.10,Bl.al.14,Bl.al2.14} Figure \ref{fig:z} shows the coordinate-invariant relationship $u^t_\text{GSF}(r_\Omega)$, as computed exactly (up to a small, controllable numerical error) in BHP theory and analytically in the PN approximation up to 4PN order.

\begin{table}[h]
	\renewcommand{\arraystretch}{1.6}
	\tbl{The analytically determined PN coefficients $\alpha_n$, $\beta_n$ and $\gamma_n$ entering the weak-field expansion \eqref{ut_GSF} of the conservative GSF contribution to the redshift observable $u^t = 1/z$, up to 7.5PN order. Linear and quadratic logarithmic runnings appear at leading 4PN and 7PN orders, while half-integral contributions appear at leading 5.5PN order. The coefficients in bold font were tested against independent PN calculations valid for arbitrary mass ratios.}
	{\begin{tabular}{@{}lccc@{}}
		\toprule
		$n$ & $\alpha_n$ & $\beta_n$ & $\gamma_n$ \\
		\colrule
		0 & $\boldsymbol{-1}$ & \\
		1 & $\boldsymbol{-2}$ & \\
		2 & $\boldsymbol{-5}$ & \\
		3 & $\boldsymbol{- \frac{121}{3} + \frac{41}{32} \pi^2}$ & \\
		4 & $\boldsymbol{- \frac{1157}{15} + \frac{677}{512} \pi^2 - \frac{128}{5} \gamma_\textbf{E} - \frac{256}{5} \ln{2}}$ & $\boldsymbol{-\frac{64}{5}}$ \vspace{0.1cm} \\
		5 & \parbox{3.8cm}{$\frac{1606877}{3150} - \frac{60343}{768} \pi^2 + \frac{1912}{105} \gamma_\text{E}$ \vspace{0.1cm} \\ $\text{    } + \frac{7544}{105} \ln{2} - \frac{243}{7} \ln{3}$} & $\boldsymbol{+\frac{956}{105}}$ \\
		5.5 & $\boldsymbol{-\frac{13696}{525} \pi}$ & \vspace{0.1cm} \\
		6 & \parbox{5cm}{$\frac{17083661}{4050} - \frac{1246056911}{1769472} \pi^2 + \frac{2800873}{262144} \pi^4$ \vspace{0.1cm} \\ $\text{  } + \frac{102512}{567} \gamma_\text{E} + \frac{372784}{2835} \ln{2} + \frac{1215}{7} \ln{3}$} & $+\frac{51256}{567}$ \\
		6.5 & $\boldsymbol{+\frac{81077}{3675} \pi}$ & \vspace{0.1cm} \\
		7 & \parbox{6.5cm}{$\frac{12624956532163}{382016250} - \frac{9041721471697}{2477260800} \pi^2 - \frac{23851025}{16777216} \pi^4$ \vspace{0.1cm} \\ $- \frac{10327445038}{5457375} \gamma_\text{E} - \frac{16983588526}{5457375} \ln{2} - \frac{2873961}{24640} \ln{3}$ \vspace{0.1cm} \\ $\text{  } \hspace{0.2cm} - \frac{1953125}{19008} \ln{5} + \frac{109568}{525} \gamma_\text{E}^2 + \frac{438272}{525} \gamma_\text{E} \ln{2}$ \vspace{0.1cm} \\ $\text{  } \hspace{1.2cm} + \frac{438272}{525} \ln^2{(2)} - \frac{2048}{5} \zeta(3)$} & \parbox{3cm}{$\frac{5163722519}{5457375} - \frac{109568}{525} \gamma_\text{E}$ \vspace{0.1cm} \\ $\text{    } \hspace{0.4cm} - \frac{219136}{525} \ln{2}$} & $+\frac{27392}{525}$ \\
		7.5 & $\boldsymbol{+\frac{82561159}{467775} \pi}$ & \\
		\botrule
	\end{tabular}\label{tab:coeffs}}
\end{table}

\section{Geodetic Spin Precession}\label{sec:psi}

Now, endow the particle with a spin $s_a$ whose magnitude is taken to be small enough so as not to affect its motion. Such a particle follows a timelike geodesic $\gamma$ with unit tangent $u^a$, with its spin parallel-transported along that geodesic:
\beq\label{du-ds}
	u^b \nabla_b u^a = 0 \, , \qquad u^b \nabla_b s_a = 0 \, .
\eeq
It immediately follows that the magnitudes $u^a u_a = -1$ and $s^a s_a \equiv s^2$ are conserved along $\gamma$. The scalar product $u^a s_a$ is also conserved, as consistent with the physical requirement that the spin be spatial in the particle's rest frame: $u^a s_a = 0$.

Although the spin's magnitude is conserved, its direction may, however, precess. By introducing along $\gamma$ an orthonormal triad $e_i^{\phantom{i}a}$ ($i=1,2,3$) of vectors orthogonal to $u^a$, the equation of parallel transport for the spin $s_a$ can be recast into a Newtonian-looking (but exact) equation for the spin frame components ${(\bm{s})}_i \equiv e_i^{\phantom{i}a} s_a$, namely
\beq
	\frac{\ud \bm{s}}{\ud t} = \bm{\omega} \times \bm{s} \, .
\eeq
The spin precession frequency $\bm{\omega}$ does, in general, depend on the choice of frame $e_i^{\phantom{i}a}$. However, for a circular orbit, the existence of a Killing vector such that $k^a\vert_\gamma = z \, u^a$ singles out a class of frames that are ``comoving'' with the particle, in the sense that they are Lie-dragged along $k^a$. For any frame within this class, it is easily shown that both $\bm{\omega}$ and $\bm{\omega} \cdot \bm{s}$ are constant along $\gamma$. Moreover, ${(\bm{\omega})}_i = \frac{1}{2} e_i^{\phantom{i}a} \varepsilon_{abcd} \, u^b \nabla^c k^d$, where $\varepsilon_{abcd}$ is the natural volume element associated with the metric. The Euclidean norm $\omega^2 \equiv \bm{\omega} \cdot \bm{\omega}$ of the precession frequency is given by the manifestly coordinate-invariant and frame-invariant expression\cite{Do.al.14}
\beq\label{omega}
	\omega^2 = \frac{1}{2} \, \nabla^a k^b \nabla_a k_b \vert_\gamma \, .
\eeq
While the redshift observable \eqref{z} probes the helical Killing field $k^a$ along the particle's worldline, the spin precession frequency \eqref{omega} probes its gradient.\footnote{Interestingly, the expression \eqref{omega} for the norm of the spin precession frequency $\bm{\omega}$ coincides with that of the surface gravity $\kappa$ of an equilibrium black hole whose horizon is generated by $k^a$.\cite{Ba.al.73,Fr.al.02,GrLe.13}} A convenient, intuitive measure of the spin precession effect is given by the \textit{precession angle} per radian of orbital revolution, namely $\psi \equiv 1 - \omega / \Omega$.\footnote{The extra term accounts for the fact that, in addition to the precession effect, the azimuthal angle swept by the spin $\bm{s}$ accumulates an additional $2\pi$ radians for every orbital revolution.\cite{Do.al.14,BiDa3.14}} The relation $\psi(\Omega)$ is coordinate-invariant and can be computed in both the PN and GSF frameworks.

The expression for $\psi(\Omega)$ has been computed in the PN approximation up to the next-to-next-to-leading order, using two independent methods: from the knowledge of the 3PN near-zone metric in harmonic coordinates,\cite{Bo.al.13} and from that of the spin-orbit contributions to the binary canonical Hamiltonian in ADM-TT coordinates.\cite{Do.al.14} The result, which is valid for any mass ratio, reads
\begin{align}\label{psi_PN}
	\psi &= \left( \frac{3}{4} + \frac{3}{4} \Delta + \frac{\nu}{2} \right) x + \left( \frac{9}{16} + \frac{9}{16} \Delta + \frac{5}{4} \nu - \frac{5}{8} \Delta \, \nu - \frac{\nu^2}{24} \right) x^2 \nonumber \\ &+ \left( \frac{27}{32} + \frac{27}{32} \Delta + \frac{3}{16} \nu - \frac{39}{8} \Delta \, \nu - \frac{105}{32} \nu^2 + \frac{5}{32} \Delta \, \nu^2 - \frac{\nu^3}{48} \right) x^3 + o(x^3) \, .
\end{align}
Since this ``geodetic'' spin precession effect is related to spacetime curvature, $\psi \to 0$ in the weak-field limit $x \to 0$. In the test-particle limit $m_1 \to 0$, we recover the prediction $\psi \simeq \frac{3}{2} \, \Omega^{-1} \vert \bm{v} \times \bm{\nabla} \Phi \vert \simeq \frac{3}{2} \, y$ from de Sitter's formula for the leading-order precession angle of a test gyro moving with an orbital velocity $\bm{v}$ in a gravitational potential $\Phi$.\cite{deS.16}

\begin{figure}
	\begin{center}
		\includegraphics[width=0.85\linewidth]{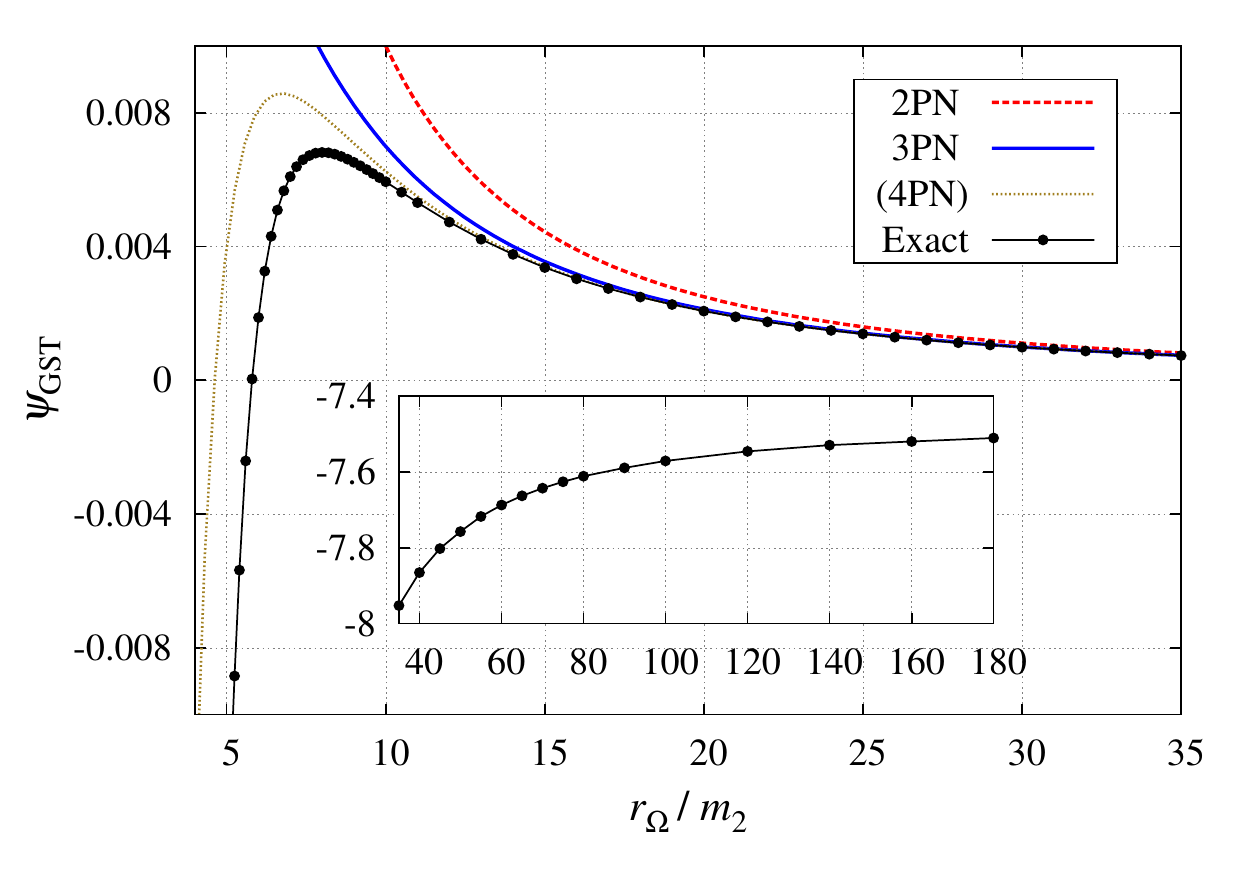}
		\caption{The conservative gravitational self-torque contribution $\psi_\text{GST}$ to the spin precession angle as a function of the orbit separation $r_\Omega$. The solid black line interpolates the numerical GST data, while the dashed red and solid blue lines show the 2PN and 3PN predictions for comparison. The inset, showing the difference between the GST and 3PN results multiplied by $(r_\Omega / m_2)^4$, hints at the value of the 4PN coefficient. The dotted brown line displays the 4PN curve obtained using the value $-15/2$ of that coefficient, as subsequently computed. \textit{Reproduced from Ref.~\protect\refcite{Do.al.14}.}}
		\label{fig:psi}
	\end{center}
\end{figure}

On the other hand, for a spinning particle of mass $m_1$ orbiting a Schwarzschild black hole of mass $m_2 \gg m_1$, the spin precession angle can be written as
\beq\label{psi_BH}
	\psi = \psi_\text{Schw}(y) + q \, \psi_\text{GST}(y) + \calO(q^2) \, ,
\eeq
where the leading contribution is known analyticaly as $\psi_\text{Schw}(y) = 1 - \sqrt{1-3y}$.\cite{Str,Do.al.14} The limit $\nu \to 0$ (and $\Delta \to 1$) of the 3PN result \eqref{psi_PN} is in agreement with the 3PN expansion of this test-mass result. The $\calO(q)$ correction to the geodesic precession rate caused by the backreaction of the conservative piece of the body's gravitational field may be interpreted as a gravitational ``self-torque''(GST) effect. Reference \refcite{Do.al.14} computed the invariant function $\psi_\text{GST}(\Omega)$ numerically, with high accuracy, for a range of separations $4 m_2 \leqslant r_\Omega \leqslant 180 m_2$. In the weak-field regime $y \ll 1$, the 3PN result \eqref{psi_PN} yields $\psi_\text{GST}(y) = y^2 - 3 y^3 + o(y^3)$, showing that the $\calO(q)$ correction contributes at leading 2PN order. Figure \ref{fig:psi} displays the numerical results for $\psi_\text{GST}(r_\Omega)$, together with the 2PN and 3PN approximations. The inset, showing the difference between the GST and 3PN results multiplied by $(r_\Omega / m_2)^4$, hints at the value of the 4PN coefficient. The dotted brown line displays the 4PN curve using the value $-15/2$ of that coefficient, as subsequently computed in Ref.~\refcite{BiDa3.14}, which used perturbative techniques valid to linear order in the mass ratio to analytically compute the weak-field expansion of $\psi_\text{GST}(y)$ up to 8.5PN order.

\section{Relativistic Periastron Advance}\label{sec:K}

Next, we review an ``extensive'' comparison of the predictions from NR simulations, BHP theory and the PN approximation, based on the (Mercury-type) general relativistic periastron advance. For non-spinning black hole binaries, a generic (bound) orbit can be uniquely specified, up to initial conditions, by its binding energy and angular momentum (see Sec.~\ref{sec:E}), or alternatively using the two natural frequencies of the motion
\beq\label{freq}
	\Omega_r \equiv \frac{2\pi}{T_r} \, , \qquad \langle \Omega_\varphi \rangle \equiv \frac{1}{T_r} \int_0^{T_r} \! \dot{\varphi}(t) \, \ud t \, ,
\eeq
where $T_r$ is the radial period (from periastron to periastron).\footnote{Actually, for a test particle orbiting a Schwarzschild black hole, there are two physically distinct bound orbits corresponding to any given pair of frequencies $\{ \Omega_r,\langle \Omega_\varphi \rangle \}$, one of them lying deep in the zoom-whirl regime.\cite{BaSa.11,Wa.al.13}} In the contexts of the PN approximation and BHP theory, which allow for a point-particle description of (at least one of the) black holes, the frequencies \eqref{freq} can be shown to be invariant under a large class of ``physically reasonable'' coordinate transformations.\cite{DaSc.88,Me.al2.04,BaSa.11} The ratio of these frequencies,
\beq
	K \equiv \frac{\langle \Omega_\varphi \rangle}{\Omega_r} = 1 + \frac{\Delta \Phi}{2\pi} \, ,
\eeq
is related in a simple way to the angular advance of the periastron per radial period, $\Delta \Phi$. In Newtonian gravitation, eccentric orbits or closed: $\Delta \Phi = 0$. In Einsteinian gravitation, they are not: $\Delta \Phi > 0$. One of the earliest successes of general relativity was to correctly account for the observed anomalous advance of Mercury's perihelion $\sim 43''/ \text{cent.}$ through the leading-order (1PN) relativistic angular advance, $\Delta \Phi = 6\pi M_\odot / [a(1-e^2)]$, where $M_\odot$ is the mass of the Sun, while $a$ and $e$ are the semi-major axis and eccentricity of Mercury's orbit, respectively.

In the limit of vanishing eccentricity, $e \to 0$, $\langle \Omega_\varphi \rangle$ reduces to the circular-orbit frequency $\Omega$ introduced earlier, and the relationship $K(\Omega)$ is coordinate-invariant.\cite{BaSa.11} For non-spinning compact binaries, it has been computed up to 3PN order.\cite{Ro.38,DaSc.88,Da.al.00} The result, which is valid for any mass ratio, reads
\begin{align}\label{K_PN}
	K = 1 &+ 3 x + \bigg( \frac{27}{2} - 7 \nu \biggr) \, x^2 \nonumber \\ &+ \biggl( \frac{135}{2} - \biggl[ \frac{649}{4} - \frac{123}{32} \pi^2 \biggr] \nu + 7 \nu^2 \biggr) \, x^3 + o(x^3) \, .
\end{align}
Notice that finite mass-ratio corrections, \textit{i.e.} terms proportional to $\nu$ and $\nu^2$ in \eqref{K_PN}, appear at leading 2PN and 3PN orders, respectively. On the other hand, for a test particle of mass $m_1$ on a circular geodesic orbit about a non-rotating black hole of mass $m_2$, the circular-orbit periastron advance is known in closed form as \cite{Cu.al.94}
\beq\label{K_Schw}
	K_\text{Schw}(y) = \frac{1}{\sqrt{1-6y}} \, .
\eeq
This formula is singular at the Schwarzschild innermost stable circular orbit (ISCO), located at $r_\Omega = 6m_2$, because the radial frequency $\Omega_r$ vanishes there (by definition). The test-particle limit $m_1 \to 0$ of the 3PN result \eqref{K_PN} agrees with the 3PN expansion of the test-mass result \eqref{K_Schw}. The conservative $\calO(q)$ correction to this leading-order result, say $q \, K_\text{GSF}(y)$, was computed numerically in Refs.~\refcite{Ba.al.10} and \refcite{BaSa.11}, for orbital separations in the range $6 m_2 < r_\Omega \leqslant 80 m_2$. In the weak-field limit $y \to 0$, this GSF data was found to be in good agreement with the PN expansion
\begin{align}
	K_\text{GSF}(y) &= 2 y + 11 y^2 - \left( \frac{109}{4} - \frac{123}{32} \pi^2 \right) y^3 \nonumber \\ &+ \left( k_4 - \frac{1256}{15} \ln{y} \right) y^4 + \left( k_5 + \frac{1964}{35} \ln{y} \right) y^5 + o(y^5) \, ,
\end{align}
as derived from \eqref{K_PN} for the 1PN, 2PN and 3PN terms, and in Ref.~\refcite{Ba.al.10} for the leading and next-to-leading logarithmic contributions at 4PN and 5PN orders. Moreover, the GSF data was used to estimate the numerical values of the 4PN and 5PN polynomial coefficients $k_4$ and $k_5$, as well as the sign of the 6PN logarithmic coefficient.\cite{Ba.al.10}

Now, motivated by the ``mathematical structure'' of the PN formula \eqref{K_PN}, which clearly accounts for the \textit{discrete symmetry} by exchange $1 \leftrightarrow 2$ of the bodies' labels, and using the perturbative equalities $q = \nu + \calO(\nu^2)$ and $y = x - \frac{2}{3} \, \nu \, x + \calO(\nu^2)$, we may rewrite the perturbative result for $K = K_\text{Schw}(y) + q \, K_\text{GSF}(y) + \calO(q^2)$ in the alternative, symmetric form
\beq\label{K_BH}
	K = K_\text{Schw}(x) + \nu \, \tilde{K}_\text{GSF}(x) + O(\nu^2) \, ,
\eeq
where $\tilde{K}_\text{GSF}(x) = K_\text{GSF}(x) - \frac{2}{3} \, x \, K'_\text{Schw}(x)$ encodes the leading effects of finite mass-ratio corrections. Equation \eqref{K_BH} can be interpreted as a formal PN series of the type \eqref{K_PN} that would include \textit{all} PN contributions at $\calO(\nu^0)$ and $\calO(\nu)$.

Following the breakthrough in the numerical simulation of the late inspiral and merger of black hole binaries, it has recently become possible to study the periastron advance in full general relativity. The relationship $K(\Omega)$ was first extracted from NR simulations of non-spinning black hole binaries moving on quasi-circular orbits in Ref.~\refcite{Mr.al.10}. An improved analysis that made use of longer and more accurate numerical simulations was performed in Ref.~\refcite{Le.al.11}, where the periastron advance of non-spinning binaries with mass ratios $q = 1$, $2/3$, $1/3$, $1/5$, $1/6$ and $1/8$ was measured in the frequency range $0.01 \lesssim M\Omega \lesssim 0.035$, with a relative uncertainty $\sim 0.1\%-1\%$. For the range of inspiral orbits covered by these NR simulations, $0.3\% \lesssim \dot{\Omega} / \Omega^2 \lesssim 1.7\%$, ensuring that the evolution is adiabatic (recall Sec.~\ref{sec:h}). Hence, these numerical results could be used to assess the performance of several (semi-)analytical models that rely on the adiabatic approximation, including the 3PN result \eqref{K_PN} and the perturbative expansion \eqref{K_BH}.

\begin{figure}
	\begin{center}
		\includegraphics[width=\linewidth]{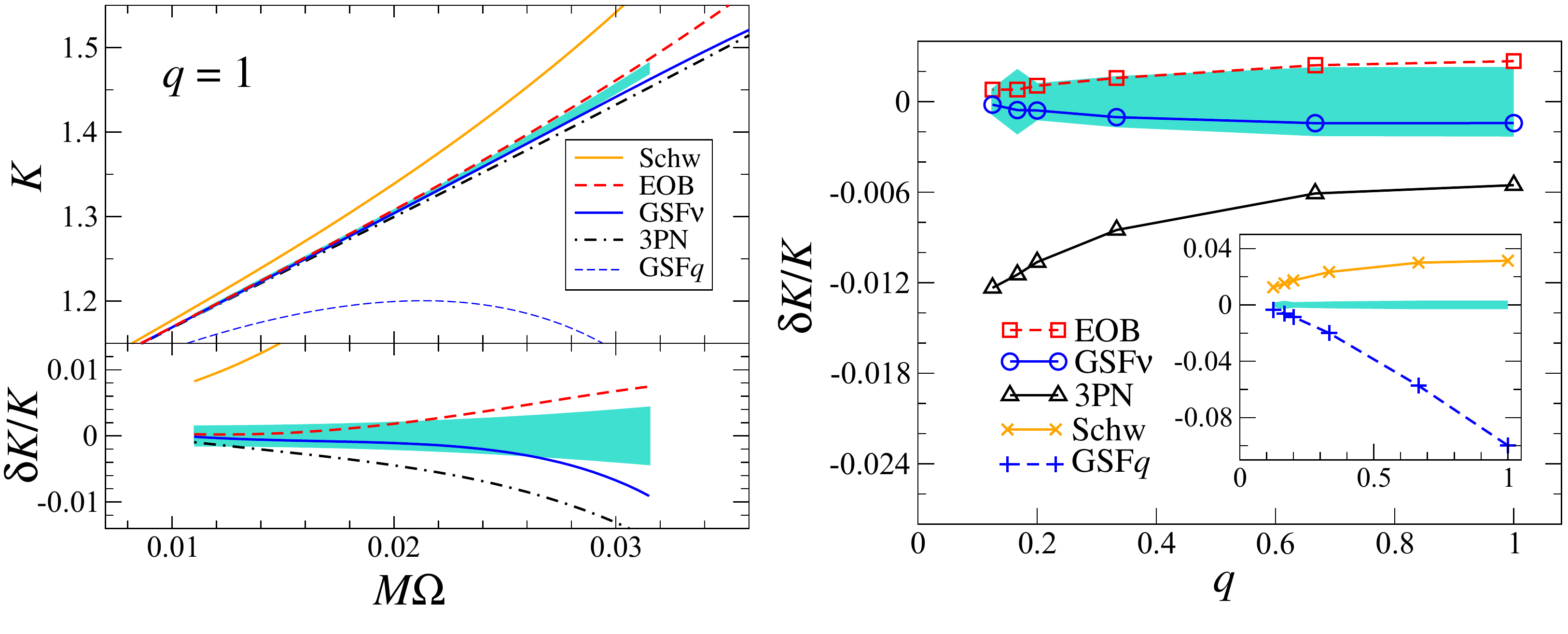}
		\caption{The periastron advance $K$ of an equal-mass, non-spinning black hole binary, in the limit of vanishing eccentricity, as a function of the circular-orbit frequency $\Omega$ (left panel), and the relative difference $\delta K / K \equiv K / K_\text{NR} - 1$ as a function of the mass ratio $q$, for $M\Omega = 0.022$ (right panel). The cyan-shaded area marks the error margin of the NR data. \textit{Reproduced from Ref.~\protect\refcite{Le.al.11}.}}
		\label{fig:K}
	\end{center}
\end{figure}

The left panel of Fig.~\ref{fig:K} shows the various $K(\Omega)$ curves for equal-mass binaries, as computed in NR (cyan-shaded region), in PN theory (``3PN''), in the EOB model of Refs.~\refcite{Da.10} and \refcite{Da.al3.00} (``EOB''), in the test-mass approximation (``Schw''), and including the conservative self-force correction while using the usual mass ratio $q$ (``GSF$q$'') or the symmetric mass ratio $\nu$ (``GSF$\nu$''). The right panel shows the relative difference $\delta K / K \equiv K / K_\text{NR} - 1$ as a function of the mass ratio $0 < q \leqslant 1$ for $M \Omega = 0.022$. Despite the good agreement between the 3PN and NR results for equal masses, with $\lesssim 1 \%$ relative difference even at the high-frequency end, the accuracy of the 3PN formula \eqref{K_PN} deteriorates with decreasing $q$, confirming that the PN approximation performs best for comparable masses.\cite{Bl.02,Fa.11} More remarkably, while the agreement between the GSF$q$ and NR results becomes manifest only at sufficiently small $q$, as might be expected, the GSF$\nu$ prediction \eqref{K_BH} agrees very well with the NR data at \textit{all} frequencies and for \textit{all} mass ratios considered, including the equal mass case.

\section{Binding Energy and Angular Momentum}\label{sec:E}

In the context of the PN approximation, when restricting to the conservative part of the dynamics of a binary system of compact objects, the perturbative invariance of the Lagrangian (in harmonic coordinates) under the Poincar\'e group implies the existence of ten Noetherian conserved quantities.\cite{Da.al2.00,deA.al.01} In the center-of-mass frame, the only non-vanishing conserved quantities are the binding energy $E$ and the orbital angular momentum $J$. For non-spinning bodies, these have been computed up to 3PN order by a variety of groups using different formalisms, gauge conditions, and models for the compact objects.\cite{JaSc.98,BlFa2.00,Da.al2.00,Da.al.00,Da.al.01,BlFa.01,ItFu.03,Bl.al.04,It.04,FoSt.11} More recently, the 4PN contribution to the binding energy $E$ has also been computed.\cite{Bl.al2.10,FoSt.11,FoSt.13,JaSc.12,JaSc.13,Da.al.14}. For circular orbits, the 4PN-accurate expressions of the specific binding energy $\hat{E} \equiv E / \mu$ and dimensionless angular momentum $\hat{J} \equiv J / (M \mu)$ read\footnote{The 4PN contribution to the angular momentum was derived here by application of the first law of binary mechanics, which implies $\partial \hat{E} / \partial \Omega = \Omega \, \partial \hat{J} / \partial \Omega$ at fixed $m_1$ and $m_2$.\cite{Le.al.12}}

\begin{subequations}\label{EJ_PN}
	\begin{align}
		\hat{E} &= - \frac{x}{2} \, \biggl\{ 1 + \left( - \frac{3}{4} - \frac{\nu}{12} \right) x + \left( - \frac{27}{8} + \frac{19}{8} \nu - \frac{\nu^2}{24} \right) x^2 \nonumber \\ &+ \left( - \frac{675}{64} + \left[ \frac{34445}{576} - \frac{205}{96} \pi^2 \right] \nu - \frac{155}{96} \nu^2 - \frac{35}{5184} \nu^3 \right) x^3 \nonumber \\ & + \left( - \frac{3969}{128} + \left[ - \frac{123671}{5760} + \frac{9037}{1536} \pi^2 + \frac{896}{15} \gamma + \frac{448}{15} \ln{(16x)} \right] \nu \right. \nonumber \\ &\qquad \left. + \left[ - \frac{498449}{3456} + \frac{3157}{576} \pi^2 \right] \nu^2 + \frac{301}{1728} \nu^3 + \frac{77}{31104} \nu^4 \right) x^4 + o(x^4) \biggr\} \, , \\
		\hat{J} &= \frac{1}{\sqrt{x}} \, \biggl\{ 1 + \left( \frac{3}{2} + \frac{\nu}{6} \right) x + \left( \frac{27}{8} - \frac{19}{8} \nu + \frac{\nu^2}{24} \right) x^2 \nonumber \\ &+ \left( \frac{135}{16} + \biggl[ - \frac{6889}{144} + \frac{41}{24} \pi^2 \biggr] \nu + \frac{31}{24} \nu^2 + \frac{7}{1296} \nu^3 \right) x^3 \nonumber \\ & + \left( \frac{2835}{128} + \left[ \frac{98869}{5760} - \frac{6455}{1536} \pi^2 - \frac{128}{3} \gamma - \frac{64}{3} \ln{(16x)} \right] \nu \right. \nonumber \\ &\qquad \left. + \left[ \frac{356035}{3456} - \frac{2255}{576} \pi^2 \right] \nu^2 - \frac{215}{1728} \nu^3 - \frac{55}{31104} \nu^4 \right) x^4 + o(x^4) \biggr\} \, .
	\end{align}
\end{subequations}

On the other hand, in the extreme mass-ratio limit, the well-known expressions of the conserved specific binding energy and dimensionless orbital angular momentum of a test mass $m_1$ moving on a circular geodesic orbit of angular frequency $\Omega$ around a Schwarzschild black hole of mass $m_2$ read
\beq\label{EJ_Schw}
	e_\text{Schw}(y) = \frac{1-2y}{\sqrt{1-3y}} - 1 \, , \quad j_\text{Schw}(y) = \frac{1}{\sqrt{y(1-3y)}} \, .
\eeq
Both expressions are singular at the Schwarzschild lightring. The test-particle limit $m_1 \to 0$ of the PN formulas \eqref{EJ_PN} agree with the 4PN expansions of the test-particle results \eqref{EJ_Schw}. Making use of the first law of binary black hole mechanics,\cite{Le.al.12} together with GSF results for the redshift observable $z(\Omega)$ (see Sec.~\ref{sec:z}), Ref.~\refcite{Le.al2.12} could compute the conservative $\calO(q)$ corrections, say $q \, e_\text{GSF}(y)$ and $q \, j_\text{GSF}(y)$, to the leading-order results \eqref{EJ_Schw}. Then, using the equalities $q = \nu + \calO(\nu^2)$ and $y = x - \frac{2}{3} \, \nu \, x + \calO(\nu^2)$, the prediction from linear perturbation theory for the relationship $\hat{E}(\hat{J})$ can be written in a parametric form, to linear order in the symmetric mass ratio, as
\begin{subequations}\label{EJ_BH}
	\begin{align}
		\hat{E} &= e_\text{Schw}(x) + \nu \, \tilde{e}_\text{GSF}(x) + O(\nu^2) \, , \\
		\hat{J} &= j_\text{Schw}(x) + \nu \, \tilde{j}_\text{GSF}(x) + O(\nu^2) \, ,
	\end{align}
\end{subequations}
where $\tilde{e}_\text{GSF}(x) = e_\text{GSF}(x) - \frac{2}{3} \, x \, e'_\text{Schw}(x)$ and similarly for $\tilde{j}_\text{GSF}(x)$.

Finally, in the context of NR simulations of inspiralling compact-object binaries, the binding energy is defined by $E(u) \equiv M_\text{B}(u) - M$, where $M = m_1 + m_2$ is the sum of the irreducible masses of the black hole's apparent horizons, while the Bondi mass $M_\text{B}(u)$ at the retarded time $u$ is computed by subtracting from the (constant) ADM mass $M_\text{ADM}$ the integrated flux of energy $\calF(u')$ carried away by gravitational radiation, \textit{i.e.} as $M_\text{B}(u) = M_\text{ADM} - \int_{-\infty}^u \calF(u') \, \ud u'$.\cite{AsMa.79} The total angular momentum $J(u)$ is obtained in a similar manner. Using accurate NR simulations,\cite{Po.al3.11} Ref.~\refcite{Da.al.12} computed the invariant relationship $\hat{E}(\hat{J})$ for non-spinning black hole binaries with mass ratios $q=1$, $1/2$ and $1/3$, and compared their results to the predictions from several ``flavors'' of the EOB model. Reference \refcite{Le.al2.12} then used these numerical results to assess the performance of the predictions from PN theory at 3PN order and linear BHP theory, as given in parametric form by \eqref{EJ_PN} [up to relative $\calO(x^3)$] and \eqref{EJ_BH}.

\begin{figure}
	\begin{center}
		\includegraphics[width=0.57\linewidth]{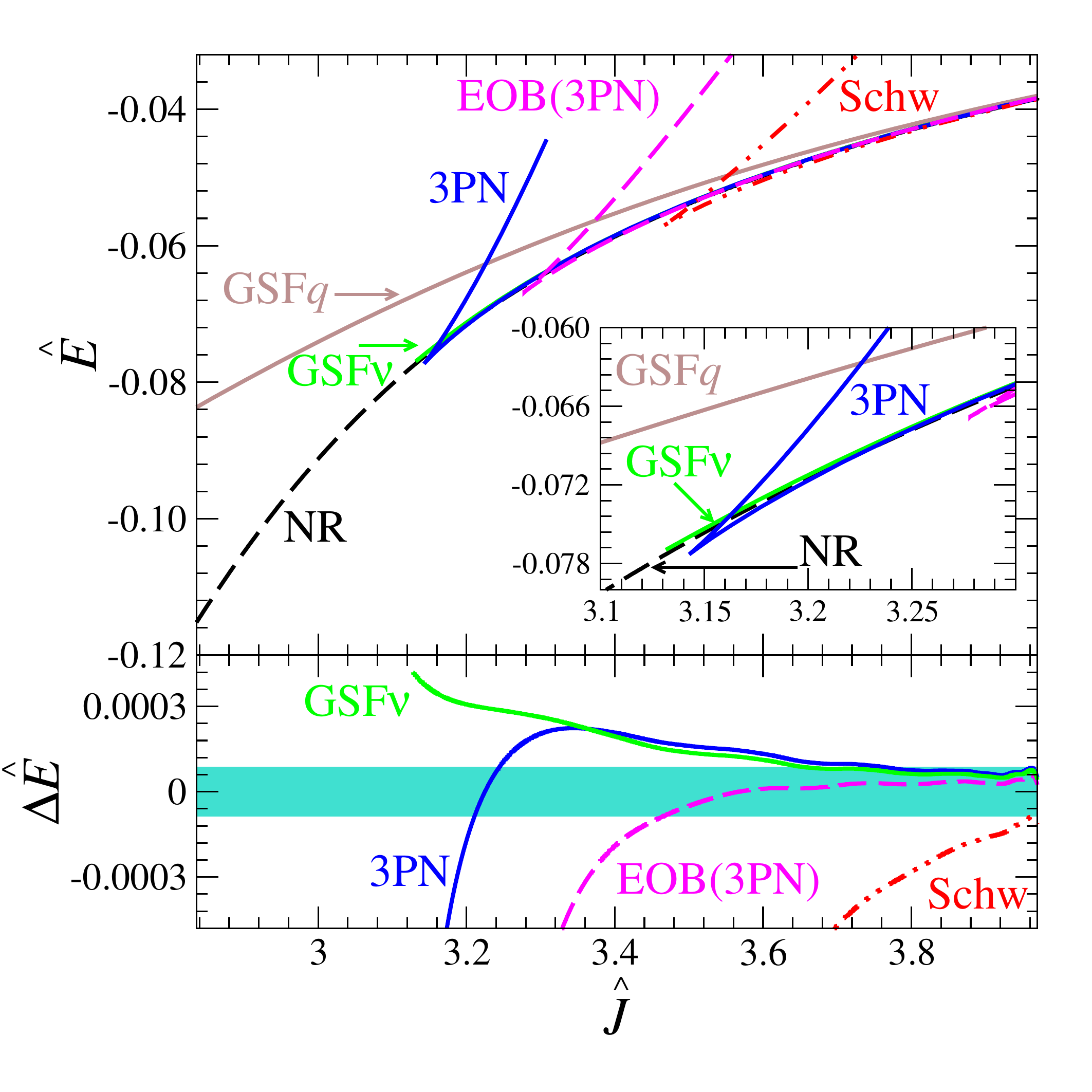}
		\caption{The specific binding energy $\hat{E} = E / \mu$ of an equal-mass, non-spinning black hole binary as a function of the dimensionless angular momentum $\hat{J} = J / (M \mu)$, as computed in numerical relativity (NR), in PN theory (3PN), in the EOB model [EOB(3PN)], in the test-mass approximation (Schw), and including the conservative gravitational self-force (GSF$q$ and GSF$\nu$). The 3PN, EOB, and test-mass curves show cusps at their respective ISCO; the lower branches correspond to stable circular orbits, while the upper branches correspond to unstable circular orbits. \textit{Reproduced from Ref.~\protect\refcite{Le.al2.12}}.}
		\label{fig:EJ}
	\end{center}
\end{figure}

Figure \ref{fig:EJ} shows the various $\hat{E}(\hat{J})$ curves for equal-mass, non-spinning black holes binaries, as computed in NR (dashed black), in the PN approximation (solid blue), in the EOB adiabatic model of Ref.~\refcite{Da.al3.00} (dashed magenta), in the test-particle approximation (dashed-dotted red), and including the conservative GSF corrections while using the mass ratio $q$ (solid gray) or the symmetric mass ratio $\nu$ (solid green). In particular, the prediction from linear BHP theory with $q \to \nu$ is in very good agreement with the exact results from NR simulations, with a difference that grows larger than the numerical error only near $x = 1/5$. We emphasize that the NR curve was obtained from an actual binary evolution,\cite{Da.al.12} and therefore includes nonadiabatic effects during the late inspiral and plunge phases. These effects are not captured by the adiabatic (semi-)analytical models, which may in part explain the differences from the NR result at small $J$, \textit{i.e.}, at large $\Omega$.

\section{Perturbation Theory for Comparable Masses}\label{sec:BHPq1}

Most quantities of physical interest in the binary black hole problem are ``properties'' of the \textit{binary} system. As such, they must be symmetric by exchange $1 \leftrightarrow 2$ of the bodies' labels. Explicit PN results for such quantities clearly embody this discrete symmetry. Indeed, those are typically written as expansions in powers of the small parameter $x = (M \Omega)^{2/3}$, with coefficients given by \textit{polynomials} in the symmetric mass ratio $\nu = m_1 m_2 / M^2$ [recall Eqs.~\eqref{h_22}, \eqref{F_PN}, \eqref{K_PN} and \eqref{EJ_PN}].\footnote{As properties of the individual constituants of the binary system, the redshift observable and the spin precession angle are not symmetric, such that the PN expansions \eqref{z_PN} and \eqref{psi_PN} depend also on the reduced mass difference $\Delta = (m_2-m_1)/M$, which changes sign under exchange $m_1 \leftrightarrow m_2$.} By contrast, in BHP theory this discrete symmetry property is broken, as one expands all quantities in powers of the small (asymmetric) mass ratio $q = m_1 / m_2$, with coefficients depending on $y = (m_2 \Omega)^{2/3}$ [see Eqs.~\eqref{F_BH}, \eqref{z_BH} and \eqref{psi_BH}]. However, as discussed in Secs.~\ref{sec:h}, \ref{sec:F}, \ref{sec:K} and \ref{sec:E}, the symmetry by exchange $1 \leftrightarrow 2$ of the bodies' labels can easily be restored in the perturbative results, yielding expansions of the type\footnote{We restrict the discussion to non-spinning black hole binaries moving along quasi-circular orbits. A possible generalization to (non-precessing) spinning binaries is discussed in Ref.~\refcite{Le.al.13}.}
\beq\label{F}
	f(\Omega;m_1,m_2) = \nu^p \sum_{n \geqslant 0} \nu^n f_{(n)}(x) \, ,
\eeq
where $p$ is a non-negative integer; for instance $p = 0$ for the periastron advance, $p=1$ for the wave polarizations, the binding energy and the orbital angular momentum, and $p=2$ for the gravitational luminosity [see, \textit{e.g.}, Eqs.~\eqref{K_BH} and \eqref{EJ_BH}].

As illustrated by Figs.~\ref{fig:psi_q10}, \ref{fig:K} and \ref{fig:EJ}, expansions of the type \eqref{F} perform remarkably well when compared to the exact results from NR simulations of \textit{comparable-mass} black hole binaries, even when applied at leading ($n=0$) or next-to-leading ($n \leqslant 1$) order. This astonishing finding can be understood, at a heuristic level, as follows: first, in the formal expansion \eqref{F}, the finite mass-ratio corrections $\nu^n f_{(n)}(x)$ with $n \geqslant 2$ are suppressed by factors of $\nu^n$ and $\nu^{n-1}$ relative to the leading contributions $f_{(0)}(x)$ and $\nu f_{(1)}(x)$, where the symmetric mass ratio ranges in $0 < \nu \leqslant 1/4$. Second, the term $\calO(\nu^{p+n}$) in Eq.~\eqref{F} contributes at the leading $n$PN (or higher) order, \textit{i.e.}, finite mass-ratio corrections are further suppressed by increasingly higher powers of $0 < x \lesssim 1/6$. This observation suggests that BHP theory may find applications in a broader range of physical problems than previously thought, including the radiative inspiral of intermediate mass-ratio and comparable-mass binaries.

\begin{figure}
	\begin{center}
		\includegraphics[width=\linewidth]{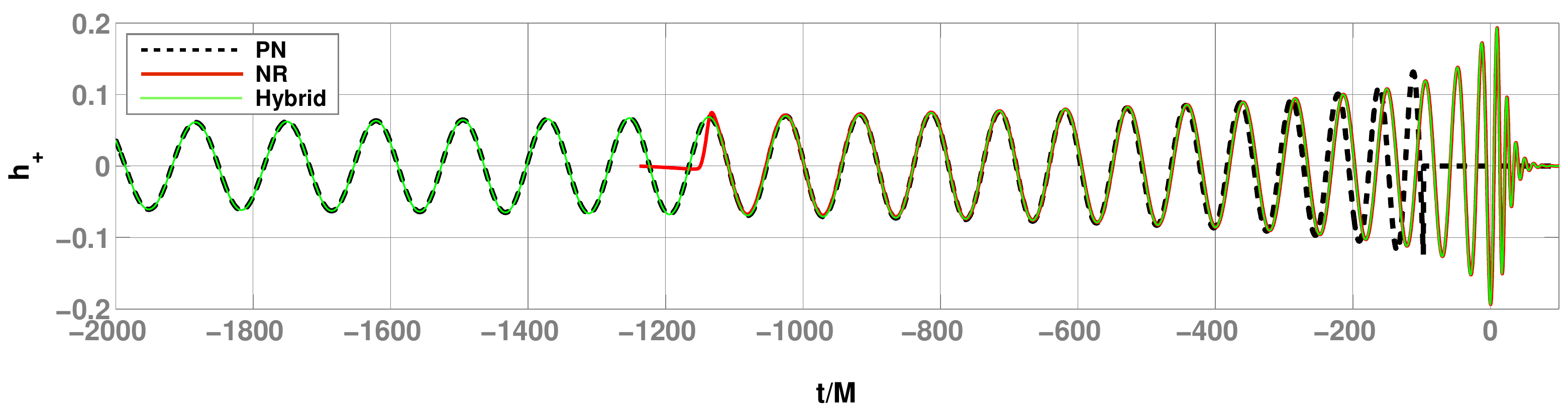}
		\caption{The polarization $h_+(t)$ generated by the quasi-circular inspiral and merger of an optimally oriented non-spinning binary black hole system with mass ratio $q \simeq 0.6$, as computed in PN theory (dashed black) for the inspiral phase and NR simulations (solid red) for the late inspiral, merger and ringdown phases. These are combined to construct a hybrid waveform (solid green) covering the entire binary evolution. \textit{Reproduced from Ref.~\protect\refcite{Aj.al.08}}.}
		\label{fig:hybrid}
	\end{center}
\end{figure}

Presently, one strategy to construct templates to be used for the detection and analysis of the gravitational-wave signals from comparable-mass black hole binaries is to generate a small number of \textit{hybrid waveforms} by ``stitching'' PN waveforms for the early adiabatic inspiral to a set of NR waveforms for the last orbits, plunge, merger and final ringdown; these hybrid waveforms are then used to create a bank of phenomenological inspiral-merger-ringdown templates covering the entire parameter space.\cite{Aj.al.07,Aj.al.08,Sa.al.10,Aj.al.11} (Another strategy is to use existing NR-calibrated EOB waveforms.\cite{Hi.al.14}) Figure \ref{fig:hybrid} shows an example in the case of the quasi-circular inspiral and merger of two non-spinning black holes with mass ratio $q \simeq 0.6$.

The modelling error in such hybrid PN/NR waveforms is dominated by the uncertainties associated with uncontrolled higher-order PN effects that become sizable during the late inspiral.\cite{Ha.al2.10,Bo.11,Da.al.11,Ma.al.11,Ma.al2.13}. Unfortunately, even for moderate signal-to-noise ratios ($\text{SNR} \sim 10$), demanding a modelling uncertainty (for each waveform) indistinguishable by upcoming advanced ground-based detectors would require hundreds of NR orbits, a task far out of reach of the most powerful supercomputers. Over a significant portion of the space of expected source parameters, using such template waveforms for parameter estimation purposes would yield \textit{systematic biases} on the measurement of the masses and spins that are larger than the statistical errors.\cite{Oh.12} Moreover, future space-based observatories such as the proposed eLISA mission are expected to detect gravitational radiation from supermassive black hole binaries up to redshifts $z \sim 20$, with typical $\text{SNR} \gtrsim 10-100$.\cite{Am.al.13} This makes the problem even worse, since higher SNRs put more stringent accuracy requirements to ensure that systematic errors do not dominate over statistical uncertainties.

A solution could be to extend existing PN results to account for these necessary relativistic corrections at 4PN and higher orders. Alternatively, the outcome of some of the comparisons reviewed above suggest a new strategy: build phenomenological inspiral-merger-ringdown templates based on \textit{BHP/NR hybrid waveforms}.\footnote{The known PN contributions $\calO(\nu^2)$ and $\calO(\nu^3)$, which contribute at leading 2PN and 3PN orders, respectively, could also be included in such templates if necessary.} Indeed, while the PN approximation necessarily breaks down in the strong-field regime, the predictions from perturbation theory remain valid there (recall Fig.~\ref{fig:methods}). Furthermore, the results of Refs.~\refcite{Sp.al2.11,Na.13,Le.al.11,Le.al2.12} and \refcite{Le.al.13} strongly suggest that the domain of validity of BHP theory can be pushed beyond the extreme mass-ratio limit $q \ll 1$, as long as the symmetry by exchange $1 \leftrightarrow 2$ of the bodies' labels is restored in the perturbative results. This opens up the prospect of building a new class of \textit{universal} template waveforms that could be used to model the gravitational-wave emission not only from extreme mass-ratio inspirals ($q \lesssim 10^{-4}$), but also from intermediate mass-ratio inspirals ($10^{-4} \lesssim q \lesssim 10^{-2}$) and comparable-mass compact-object binaries ($10^{-2} \lesssim q \leqslant 1$); all of them are highly promising sources for existing and planned ground-based detectors, as well as for future space-based observatories.\cite{Am.al.07,Sa.al.12,Aa.al.13,Am.al.13,BiPa.13}

\section{Summary and Prospects}

We have reviewed a sample of the large and growing body of work at the interface between numerical relativity, black hole perturbation theory, and the post-Newtonian approximation in the binary black hole problem. In particular, we emphasized the importance of using coordinate-invariant relationships to perform meaningful comparisons of the predictions from these approximation methods and numerical techniques. We have seen how such ``cross-cultural'' comparisons provide key checks of the validity of the various calculations, thereby increasing our confidence in the template waveforms that are being used to search for gravitational-wave signals from binary black holes. We have also seen how these comparisons help to delineate the domain of validity of approximation methods such as post-Newtonian expansions and black hole perturbations. In particular, we have highlighted several comparisons suggesting that perturbation theory may prove useful to model not only extreme mass-ratio inspirals, but also intermediate mass-ratio inspirals and comparable-mass binaries. This striking observation suggests an original strategy to devise a new class of universal template waveforms for black hole binaries, as discussed in Sec.~\ref{sec:BHPq1}.

An important question that has barely been addressed so far, but which deserves closer scrutiny, is that of the identification of the black holes' physical parameters. While comparing the results from PN expansions, black hole perturbations and NR simulations, it is implicitly assumed that the various notions of mass and spin used in these analytical and numerical schemes can be safely identified. Yet, the validity of this assumption is far from obvious. For instance, in NR simulations of black hole binaries, the irreducible mass and spin of each black hole are computed as surface integrals over the apparent horizon,\cite{Lo.al.08} whereas in most PN treatments of compact-object binaries, the mass and spin appear as multipole moments constructed from the body's stress-energy tensor.\cite{StPu.10}

Nevertheless, while most comparisons performed to date have been restricted to the simplest case of non-spinning black hole binaries moving along quasi-circular orbits (recall Tables \ref{tab:waveform} and \ref{tab:dynamics}), future work will most certainly focus on more generic binary configurations (\textit{e.g.} spinning back holes and precessing orbits). In particular, recent progress in perturbative GSF calculations will soon enable new comparisons with the predictions from PN theory based on (i) the redshift observable, the spin precession frequency and higher-order invariants for circular equatorial orbits in a Kerr background,\cite{Do.al.14} (ii) an averaged version of the redshift observable for eccentric orbits in a Schwarzschild background,\cite{BaSa.11} and (iii) both conservative and dissipative second-order GSF effects.\cite{Po.14} Meanwhile, PN calculations are successfully being carried up to 4PN order,\cite{JaSc.13,FoSt.13,Da.al.14} and NR simulations are being pushed towards smaller mass ratios\cite{LoZl.11} and larger separations.\cite{LoZl.13}

In summary, as numerical relativity simulations become more accurate and further extend into the corners of the parameter space, while post-Newtonian expansions are pushed to increasingly higher orders, with black hole perturbation theory tackling more generic configurations and second-order calculations, their overlap in the binary black hole problem increases steadily. Following numerous decades of independent work and painstaking progress, analytical and numerical relativists can, at last, ask the same questions and obtain consistent answers while comparing their results. Almost one century since the inception of Einstein's theory of space, time and gravitation, the general relativistic two-body problem is more alive than ever.

\section*{Acknowledgments}

It is a pleasure to thank P. Ajith, L. Barack, E. Barausse, L. Blanchet, M. Boyle, D. A. Brown, A. Buonanno, V. Cardoso, G. B. Cook, S. Detweiler, S. R. Dolan, L. E. Kidder, A. H. Mrou{\'e}, A. Nagar, H. P. Pfeiffer, N. Sago, M. A. Scheel, U. Sperhake, A. Taracchini, S. A. Teukolsky, N. Warburton and B. F. Whiting for allowing me to reproduce previously published figures. This work was supported by a Marie Curie FP7 Integration Grant within the 7th European Union Framework Programme.

\bibliographystyle{ws-ijmpd}
\bibliography{}

\begin{thebibliography}{100}

\bibitem{Da.87}
T.~Damour, { The problem of motion in {N}ewtonian and {E}insteinian gravity},
  in {\em Three hundred years of gravitation\/},  eds. S.~W. Hawking and
  W.~Israel (Cambridge University Press, Cambridge, 1987), p. 128.

\bibitem{Pr.09}
F.~Pretorius, { Binary black hole coalescence}, in {\em Physics of relativistic
  objects in compact binaries: From birth to coalescence\/},  eds. M.~Colpi,
  P.~Casella, V.~Gorini, U.~Moschella and A.~Possenti, Astrophysics and Space
  Science Library, Vol.~359 (Springer, New York, 2009), p. 305.
\newblock \href{http://arxiv.org/abs/0710.1338}{{\ttfamily arXiv:0710.1338
  [gr-qc]}}.

\bibitem{SaSc.09}
B.~S. Sathyaprakash and B.~F. Schutz, {\em Living Rev. Relativity} {\bf 12}
  (2009)  ~2, \href{http://arxiv.org/abs/0903.0338}{{\ttfamily arXiv:0903.0338
  [gr-qc]}}.

\bibitem{Aa.al.13}
J.~Aasi {\em et~al.}  (2013) \href{http://arxiv.org/abs/1304.0670}{{\ttfamily
  arXiv:1304.0670 [gr-qc]}}.

\bibitem{Am.al.13}
P.~Amaro-Seoane {\em et~al.}, {\em GW Notes} {\bf 6}  (2013)  ~4,
  \href{http://arxiv.org/abs/1201.3621}{{\ttfamily arXiv:1201.3621
  [astro-ph.CO]}}.

\bibitem{Mi2.09}
M.~C. Miller, {\em Class. Quant. Grav.} {\bf 26}  (2009)   094031,
  \href{http://arxiv.org/abs/0812.3028}{{\ttfamily arXiv:0812.3028
  [astro-ph]}}.

\bibitem{AmSa.10}
P.~Amaro-Seoane and L.~Santamaria, {\em Astrophys. J.} {\bf 722}  (2010)
  1197, \href{http://arxiv.org/abs/0910.0254}{{\ttfamily arXiv:0910.0254
  [astro-ph.CO]}}.

\bibitem{Ba.al.13}
S.~Babak {\em et~al.}, {\em Phys. Rev D} {\bf 87}  (2013)   024033,
  \href{http://arxiv.org/abs/1208.3491}{{\ttfamily arXiv:1208.3491 [gr-qc]}}.

\bibitem{Bl.al.98}
L.~Blanchet, G.~Faye and B.~Ponsot, {\em Phys. Rev. D} {\bf 58}  (1998)
  124002, \href{http://arxiv.org/abs/arXiv:gr-qc/9804079}{{\ttfamily
  arXiv:gr-qc/9804079}}.

\bibitem{BlFa.01}
L.~Blanchet and G.~Faye, {\em Phys. Rev. D} {\bf 63}  (2001)   062005,
  \href{http://arxiv.org/abs/arXiv:gr-qc/0007051}{{\ttfamily
  arXiv:gr-qc/0007051}}.

\bibitem{deA.al.01}
V.~C. de~Andrade, L.~Blanchet and G.~Faye, {\em Class. Quant. Grav.} {\bf 18}
  (2001)   753, \href{http://arxiv.org/abs/arXiv:gr-qc/0011063}{{\ttfamily
  arXiv:gr-qc/0011063}}.

\bibitem{Bl.al.04}
L.~Blanchet, T.~Damour and G.~Esposito-Far{\`e}se, {\em Phys. Rev. D} {\bf 69}
  (2004)   124007, \href{http://arxiv.org/abs/arXiv:gr-qc/0311052}{{\ttfamily
  arXiv:gr-qc/0311052}}.

\bibitem{PaWi.02}
M.~Pati and C.~M. Will, {\em Phys. Rev. D} {\bf 65}  (2002)   104008,
  \href{http://arxiv.org/abs/arXiv:gr-qc/0201001}{{\ttfamily
  arXiv:gr-qc/0201001}}.

\bibitem{MiWi.07}
T.~Mitchell and C.~M. Will, {\em Phys. Rev. D} {\bf 75}  (2007)   124025,
  \href{http://arxiv.org/abs/0704.2243}{{\ttfamily arXiv:0704.2243 [gr-qc]}}.

\bibitem{JaSc.98}
P.~Jaranowski and G.~Sch{\"a}fer, {\em Phys. Rev. D} {\bf 57}  (1998)   7274,
  \href{http://arxiv.org/abs/arXiv:gr-qc/9712075}{{\ttfamily
  arXiv:gr-qc/9712075}}, \textit{{E}rratum:} Phys. Rev. D \textbf{63},
  029902(E) (2000).

\bibitem{Da.al2.00}
T.~Damour, P.~Jaranowski and G.~Sch{\"a}fer, {\em Phys. Rev. D} {\bf 62}
  (2000)   021501(R),
  \href{http://arxiv.org/abs/arXiv:gr-qc/0003051}{{\ttfamily
  arXiv:gr-qc/0003051}}, \textit{{E}rratum:} Phys. Rev. D \textbf{63},
  029903(E) (2000).

\bibitem{Da.al.01}
T.~Damour, P.~Jaranowski and G.~Sch{\"a}fer, {\em Phys. Lett. B} {\bf 513}
  (2001)   147, \href{http://arxiv.org/abs/arXiv:gr-qc/0105038}{{\ttfamily
  arXiv:gr-qc/0105038}}.

\bibitem{JaSc.13}
P.~Jaranowski and G.~Sch{\"a}fer, {\em Phys. Rev D} {\bf 87}  (2013)   081503,
  \href{http://arxiv.org/abs/1303.3225}{{\ttfamily arXiv:1303.3225 [gr-qc]}}.

\bibitem{Da.al.14}
T.~Damour, P.~Jaranowski and G.~Sch{\"a}fer, {\em Phys. Rev. D} {\bf 89}
  (2014)   064058, \href{http://arxiv.org/abs/1401.4548}{{\ttfamily
  arXiv:1401.4548 [gr-qc]}}.

\bibitem{It.al.01}
Y.~Itoh, T.~Futamase and H.~Asada, {\em Phys. Rev. D} {\bf 63}  (2001)
  064038, \href{http://arxiv.org/abs/arXiv:gr-qc/0101114}{{\ttfamily
  arXiv:gr-qc/0101114}}.

\bibitem{ItFu.03}
Y.~Itoh and T.~Futamase, {\em Phys. Rev. D} {\bf 68}  (2003)   121501(R),
  \href{http://arxiv.org/abs/arXiv:gr-qc/0310028}{{\ttfamily
  arXiv:gr-qc/0310028}}.

\bibitem{It.04}
Y.~Itoh, {\em Phys. Rev. D} {\bf 69}  (2004)   064018,
  \href{http://arxiv.org/abs/arXiv:gr-qc/0310029}{{\ttfamily
  arXiv:gr-qc/0310029}}.

\bibitem{GoRo.06}
W.~D. Goldberger and I.~Z. Rothstein, {\em Phys. Rev. D} {\bf 73}  (2006)
  104029, \href{http://arxiv.org/abs/arXiv:hep-th/0409156}{{\ttfamily
  arXiv:hep-th/0409156}}.

\bibitem{FoSt.11}
S.~Foffa and R.~Sturani, {\em Phys. Rev. D} {\bf 84}  (2011)   044031,
  \href{http://arxiv.org/abs/1104.1122}{{\ttfamily arXiv:1104.1122 [gr-qc]}}.

\bibitem{GaLe.12}
C.~R. Galley and A.~K. Leibovich, {\em Phys. Rev. D} {\bf 86}  (2012)   044029,
  \href{http://arxiv.org/abs/1205.3842}{{\ttfamily arXiv:1205.3842 [gr-qc]}}.

\bibitem{FoSt.13}
S.~Foffa and R.~Sturani, {\em Phys. Rev. D} {\bf 87}  (2012)   064011,
  \href{http://arxiv.org/abs/1206.7087}{{\ttfamily arXiv:1206.7087 [gr-qc]}}.

\bibitem{BlDa.86}
L.~Blanchet and T.~Damour, {\em Phil. Trans. Roy. Soc. Lond. A} {\bf 320}
  (1986)   379.

\bibitem{Bl.87}
L.~Blanchet, {\em Proc. R. Soc. Lond. A} {\bf 409}  (1987)   383.

\bibitem{Bl.98}
L.~Blanchet, {\em Class. Quant. Grav.} {\bf 15}  (1998)   1971,
  \href{http://arxiv.org/abs/arXiv:gr-qc/9801101}{{\ttfamily
  arXiv:gr-qc/9801101}}.

\bibitem{WiWi.96}
C.~M. Will and A.~G. Wiseman, {\em Phys. Rev. D} {\bf 54}  (1996)   4813,
  \href{http://arxiv.org/abs/arXiv:gr-qc/9608012}{{\ttfamily
  arXiv:gr-qc/9608012}}.

\bibitem{PaWi.00}
M.~Pati and C.~M. Will, {\em Phys. Rev. D} {\bf 62}  (2000)   124015,
  \href{http://arxiv.org/abs/arXiv:gr-qc/0007087}{{\ttfamily
  arXiv:gr-qc/0007087}}.

\bibitem{Bl.al.02}
L.~Blanchet, G.~Faye, B.~R. Iyer and B.~Joguet, {\em Phys. Rev. D} {\bf 65}
  (2002)   061501, \href{http://arxiv.org/abs/arXiv:gr-qc/0105099}{{\ttfamily
  arXiv:gr-qc/0105099}}, \textit{{E}rratum:} Phys. Rev. D \textbf{71},
  129902(E) (2005).

\bibitem{Bl.al2.02}
L.~Blanchet, B.~R. Iyer and B.~Joguet, {\em Phys. Rev. D} {\bf 65}  (2002)
  064005, \href{http://arxiv.org/abs/arXiv:gr-qc/0105098}{{\ttfamily
  arXiv:gr-qc/0105098}}, \textit{{E}rratum:} Phys. Rev. D \textbf{71},
  129903(E) (2005).

\bibitem{Bl.al2.04}
L.~Blanchet, T.~Damour, G.~Esposito-Far{\`e}se and B.~R. Iyer, {\em Phys. Rev.
  Lett.} {\bf 93}  (2004)   091101,
  \href{http://arxiv.org/abs/arXiv:gr-qc/0406012}{{\ttfamily
  arXiv:gr-qc/0406012}}.

\bibitem{Bl.al2.05}
L.~Blanchet, T.~Damour, G.~Esposito-Far{\`e}se and B.~R. Iyer, {\em Phys. Rev.
  D} {\bf 71}  (2005)   124004,
  \href{http://arxiv.org/abs/arXiv:gr-qc/0503044}{{\ttfamily
  arXiv:gr-qc/0503044}}.

\bibitem{Bl.al.96}
L.~Blanchet, B.~R. Iyer, C.~M. Will and A.~G. Wiseman, {\em Class. Quant.
  Grav.} {\bf 13}  (1996)   575,
  \href{http://arxiv.org/abs/arXiv:gr-qc/9602024}{{\ttfamily
  arXiv:gr-qc/9602024}}.

\bibitem{Ar.al.04}
K.~G. Arun, L.~Blanchet, B.~R. Iyer and M.~S.~S. Qusailah, {\em Class. Quant.
  Grav.} {\bf 21}  (2004)   3771,
  \href{http://arxiv.org/abs/arXiv:gr-qc/0404085}{{\ttfamily
  arXiv:gr-qc/0404085}}, \textit{{E}rratum:} Class. Quant. Grav. \textbf{22},
  3115 (2005).

\bibitem{Ki.al.07}
L.~E. Kidder, L.~Blanchet and B.~R. Iyer, {\em Class. Quant. Grav.} {\bf 24}
  (2007)   5307, \href{http://arxiv.org/abs/0706.0726}{{\ttfamily
  arXiv:0706.0726 [gr-qc]}}.

\bibitem{Ki.08}
L.~E. Kidder, {\em Phys. Rev. D} {\bf 77}  (2008)   044016,
  \href{http://arxiv.org/abs/0710.0614}{{\ttfamily arXiv:0710.0614 [gr-qc]}}.

\bibitem{Bl.al.08}
L.~Blanchet, G.~Faye, B.~R. Iyer and S.~Sinha, {\em Class. Quant. Grav.} {\bf
  25}  (2008)   165003, \href{http://arxiv.org/abs/0802.1249}{{\ttfamily
  arXiv:0802.1249 [gr-qc]}}, \textit{{C}orrigendum:} Class. Quant. Grav.
  \textbf{29}, 239501 (2012).

\bibitem{St.al.08}
J.~Steinhoff, S.~Hergt and G.~Sch{\"a}fer, {\em Phys. Rev. D} {\bf 77}  (2008)
   081501(R), \href{http://arxiv.org/abs/0712.1716}{{\ttfamily arXiv:0712.1716
  [gr-qc]}}.

\bibitem{St.al3.08}
J.~Steinhoff, S.~Hergt and G.~Sch{\"a}fer, {\em Phys. Rev. D} {\bf 78}  (2008)
   101503(R), \href{http://arxiv.org/abs/0809.2200}{{\ttfamily arXiv:0809.2200
  [gr-qc]}}.

\bibitem{PoRo.08}
R.~A. Porto and I.~Z. Rothstein, {\em Phys. Rev. D} {\bf 78}  (2008)   044012,
  \href{http://arxiv.org/abs/0802.0720}{{\ttfamily arXiv:0802.0720 [gr-qc]}},
  \textit{{E}rratum:} Phys. Rev. D \textbf{81}, 029904(E) (2010).

\bibitem{PoRo2.08}
R.~A. Porto and I.~Z. Rothstein, {\em Phys. Rev. D} {\bf 78}  (2008)   044013,
  \href{http://arxiv.org/abs/0804.0260}{{\ttfamily arXiv:0804.0260 [gr-qc]}},
  \textit{{E}rratum:} Phys. Rev. D \textbf{81}, 029905(E) (2010).

\bibitem{He.al.10}
S.~Hergt, J.~Steinhoff and G.~Sch{\"a}fer, {\em Class. Quant. Grav.} {\bf 27}
  (2010)   135007, \href{http://arxiv.org/abs/1002.2093}{{\ttfamily
  arXiv:1002.2093 [gr-qc]}}.

\bibitem{Le.10}
M.~Levi, {\em Phys. Rev. D} {\bf 82}  (2010)   064029,
  \href{http://arxiv.org/abs/0802.1508}{{\ttfamily arXiv:0802.1508 [gr-qc]}}.

\bibitem{Le2.10}
M.~Levi, {\em Phys. Rev. D} {\bf 82}  (2010)   104004,
  \href{http://arxiv.org/abs/1006.4139}{{\ttfamily arXiv:1006.4139 [gr-qc]}}.

\bibitem{Po.10}
R.~A. Porto, {\em Class. Quant. Grav.} {\bf 27}  (2010)   205001,
  \href{http://arxiv.org/abs/1005.5730}{{\ttfamily arXiv:1005.5730 [gr-qc]}}.

\bibitem{Po.al2.11}
R.~A. Porto, A.~Ross and I.~Z. Rothstein, {\em JCAP} {\bf 1103}  (2011)   009,
  \href{http://arxiv.org/abs/1007.1312}{{\ttfamily arXiv:1007.1312 [gr-qc]}}.

\bibitem{HaSt.11}
J.~Hartung and J.~Steinhoff, {\em Ann. Phys.} {\bf 523}  (2011)   783,
  \href{http://arxiv.org/abs/1104.3079}{{\ttfamily arXiv:1104.3079 [gr-qc]}}.

\bibitem{HaSt2.11}
J.~Hartung and J.~Steinhoff, {\em Ann. Phys.} {\bf 523}  (2011)   919,
  \href{http://arxiv.org/abs/1107.4294}{{\ttfamily arXiv:1107.4294 [gr-qc]}}.

\bibitem{Le2.12}
M.~Levi, {\em Phys. Rev. D} {\bf 85}  (2012)   064043,
  \href{http://arxiv.org/abs/1107.4322}{{\ttfamily arXiv:1107.4322 [gr-qc]}}.

\bibitem{Po.al.12}
R.~A. Porto, A.~Ross and I.~Z. Rothstein, {\em JCAP} {\bf 1209}  (2012)   028,
  \href{http://arxiv.org/abs/1203.2962}{{\ttfamily arXiv:1203.2962 [gr-qc]}}.

\bibitem{Ha.al.13}
J.~Hartung, J.~Steinhoff and G.~Sch{\"a}fer, {\em Ann. Phys.} {\bf 525}  (2013)
    359, \href{http://arxiv.org/abs/1302.6723}{{\ttfamily arXiv:1302.6723
  [gr-qc]}}.

\bibitem{Bl.al2.11}
L.~Blanchet, A.~Buonanno and G.~Faye, {\em Phys. Rev. D} {\bf 84}  (2011)
  064041, \href{http://arxiv.org/abs/1104.5659}{{\ttfamily arXiv:1104.5659
  [gr-qc]}}.

\bibitem{Bu.al.13}
A.~Buonanno, G.~Faye and T.~Hinderer, {\em Phys. Rev. D} {\bf 87}  (2013)
  044009, \href{http://arxiv.org/abs/1209.6349}{{\ttfamily arXiv:1209.6349
  [gr-qc]}}.

\bibitem{Ma.al.13}
S.~Marsat, A.~Boh{\'e}, G.~Faye and L.~Blanchet, {\em Class. Quant. Grav.} {\bf
  30}  (2013)   055007, \href{http://arxiv.org/abs/1210.4143}{{\ttfamily
  arXiv:1210.4143 [gr-qc]}}.

\bibitem{Bo.al.13}
A.~Boh{\'e}, S.~Marsat, G.~Faye and L.~Blanchet, {\em Class. Quant. Grav.} {\bf
  30}  (2013)   075017, \href{http://arxiv.org/abs/1212.5520}{{\ttfamily
  arXiv:1212.5520 [gr-qc]}}.

\bibitem{Bo.al2.13}
A.~Boh{\'e}, S.~Marsat and L.~Blanchet, {\em Class. Quant. Grav.} {\bf 30}
  (2013)   135009, \href{http://arxiv.org/abs/1303.7412}{{\ttfamily
  arXiv:1303.7412 [gr-qc]}}.

\bibitem{Ma.al.14}
S.~Marsat, A.~Boh{\'e}, L.~Blanchet and A.~Buonanno, {\em Class. Quant. Grav.}
  {\bf 31}  (2014)   025023, \href{http://arxiv.org/abs/1307.6793}{{\ttfamily
  arXiv:1307.6793 [gr-qc]}}.

\bibitem{LeSt.14}
M.~Levi and J.~Steinhoff  (2014)
  \href{http://arxiv.org/abs/1408.5762}{{\ttfamily arXiv:1408.5762 [gr-qc]}}.

\bibitem{Bl.14}
L.~Blanchet, {\em Living Rev. Relativity} {\bf 17}  (2014)  ~2,
  \href{http://arxiv.org/abs/1310.1528}{{\ttfamily arXiv:1310.1528 [gr-qc]}}.

\bibitem{Sc.11}
G.~Sch{\"a}fer, { Post-{N}ewtonian methods: Analytic results on the binary
  problem}, in {\em Mass and motion in general relativity\/},  eds.
  L.~Blanchet, A.~Spallicci and B.~Whiting, Fundamental Theories of Physics,
  Vol.~162 (Springer, New York, 2011), p. 167.
\newblock \href{http://arxiv.org/abs/0910.2857}{{\ttfamily arXiv:0910.2857
  [gr-qc]}}.

\bibitem{FuIt.07}
T.~Futamase and Y.~Itoh, {\em Living Rev. Relativity} {\bf 10}  (2007)  ~2.

\bibitem{FoSt.14}
S.~Foffa and R.~Sturani, {\em Class. Quant. Grav.} {\bf 31}  (2014)   043001,
  \href{http://arxiv.org/abs/1309.3474}{{\ttfamily arXiv:1309.3474 [gr-qc]}}.

\bibitem{Bl.11}
L.~Blanchet, { Post-{N}ewtonian theory and the two-body problem}, in {\em Mass
  and motion in general relativity\/},  eds. L.~Blanchet, A.~Spallicci and
  B.~Whiting, Fundamental Theories of Physics, Vol.~162 (Springer, New York,
  2011), p. 125.
\newblock \href{http://arxiv.org/abs/0907.3596}{{\ttfamily arXiv:0907.3596
  [gr-qc]}}.

\bibitem{Mi.al.97}
Y.~Mino, M.~Sasaki and T.~Tanaka, {\em Phys. Rev. D} {\bf 55}  (1997)   3457,
  \href{http://arxiv.org/abs/arXiv:gr-qc/9606018}{{\ttfamily
  arXiv:gr-qc/9606018}}.

\bibitem{QuWa.97}
T.~C. Quinn and R.~M. Wald, {\em Phys. Rev. D} {\bf 56}  (1997)   3381,
  \href{http://arxiv.org/abs/arXiv:gr-qc/9610053}{{\ttfamily
  arXiv:gr-qc/9610053}}.

\bibitem{GrWa.08}
S.~E. Gralla and R.~M. Wald, {\em Class. Quant. Grav.} {\bf 25}  (2008)
  205009, \href{http://arxiv.org/abs/0806.3293}{{\ttfamily arXiv:0806.3293
  [gr-qc]}}.

\bibitem{DeWh.03}
S.~Detweiler and B.~F. Whiting, {\em Phys. Rev. D} {\bf 67}  (2003)   024025,
  \href{http://arxiv.org/abs/arXiv:gr-qc/0202086}{{\ttfamily
  arXiv:gr-qc/0202086}}.

\bibitem{Po2.10}
A.~Pound, {\em Phys. Rev. D} {\bf 81}  (2010)   024023,
  \href{http://arxiv.org/abs/0907.5197}{{\ttfamily arXiv:0907.5197 [gr-qc]}}.

\bibitem{Ha.12}
A.~I. Harte, {\em Class. Quant. Grav.} {\bf 29}  (2012)   055012,
  \href{http://arxiv.org/abs/1103.0543}{{\ttfamily arXiv:1103.0543 [gr-qc]}}.

\bibitem{HiFl.08}
T.~Hinderer and {\'E}.~{\'E}. Flanagan, {\em Phys. Rev. D} {\bf 78}  (2008)
  064028, \href{http://arxiv.org/abs/0805.3337}{{\ttfamily arXiv:0805.3337
  [gr-qc]}}.

\bibitem{ReWh.57}
T.~Regge and J.~A. Wheeler, {\em Phys. Rev.} {\bf 108}  (1957)   1063.

\bibitem{Ze3.70}
F.~J. Zerilli, {\em Phys. Rev. Lett.} {\bf 24}  (1970)   737.

\bibitem{Te.72}
S.~A. Teukolsky, {\em Phys. Rev. Lett.} {\bf 29}  (1972)   1114.

\bibitem{Te.73}
S.~A. Teukolsky, {\em Astrophys. J.} {\bf 185}  (1973)   635.

\bibitem{Sh.94}
M.~Shibata, {\em Phys. Rev. D} {\bf 50}  (1994)   6297.

\bibitem{PoSa.95}
E.~Poisson and M.~Sasaki, {\em Phys Rev. D} {\bf 51}  (1995)   5753,
  \href{http://arxiv.org/abs/arXiv:gr-qc/9412027}{{\ttfamily
  arXiv:gr-qc/9412027}}.

\bibitem{Mi.al.96}
Y.~Mino, M.~Shibata and T.~Tanaka, {\em Phys. Rev. D} {\bf 53}  (1996)   622.

\bibitem{Ta.al3.96}
T.~Tanaka, Y.~Mino, M.~Sasaki and M.~Shibata, {\em Phys. Rev. D} {\bf 54}
  (1996)   3762, \href{http://arxiv.org/abs/arXiv:gr-qc/9602038}{{\ttfamily
  arXiv:gr-qc/9602038}}.

\bibitem{Mi.03}
Y.~Mino, {\em Phys. Rev. D} {\bf 67}  (2003)   084027,
  \href{http://arxiv.org/abs/arXiv:gr-qc/0302075}{{\ttfamily
  arXiv:gr-qc/0302075}}.

\bibitem{Hu.00}
S.~A. Hughes, {\em Phys. Rev. D} {\bf 61}  (2000)   084004,
  \href{http://arxiv.org/abs/arXiv:gr-qc/9910091}{{\ttfamily
  arXiv:gr-qc/9910091}}, \textit{{E}rrata:} Phys. Rev. D \textbf{63}, 049902(E)
  (2001), Phys. Rev. D \textbf{65}, 069902(E) (2002), Phys. Rev. D \textbf{67},
  089901(E) (2003), Phys. Rev. D \textbf{78}, 109902(E) (2008) \& Phys. Rev. D
  \textbf{88}, 109902(E) (2013).

\bibitem{Hu.al.05}
S.~A. Hughes, S.~Drasco, {\'E}.~{\'E}. Flanagan and J.~Franklin, {\em Phys.
  Rev. Lett.} {\bf 94}  (2005)   221101,
  \href{http://arxiv.org/abs/arXiv:gr-qc/0504015}{{\ttfamily
  arXiv:gr-qc/0504015}}.

\bibitem{Sa.al.06}
N.~Sago, T.~Tanaka, W.~Hikida, K.~Ganz and H.~Nakano, {\em Prog. Theor. Phys.}
  {\bf 115}  (2006)   873,
  \href{http://arxiv.org/abs/arXiv:gr-qc/0511151}{{\ttfamily
  arXiv:gr-qc/0511151}}.

\bibitem{De.08}
S.~Detweiler, {\em Phys. Rev. D} {\bf 77}  (2008)   124026,
  \href{http://arxiv.org/abs/0804.3529}{{\ttfamily arXiv:0804.3529 [gr-qc]}}.

\bibitem{Sa.al.08}
N.~Sago, L.~Barack and S.~Detweiler, {\em Phys. Rev. D} {\bf 78}  (2008)
  124024, \href{http://arxiv.org/abs/0810.2530}{{\ttfamily arXiv:0810.2530
  [gr-qc]}}.

\bibitem{BaSa.09}
L.~Barack and N.~Sago, {\em Phys. Rev. Lett.} {\bf 102}  (2009)   191101,
  \href{http://arxiv.org/abs/0902.0573}{{\ttfamily arXiv:0902.0573 [gr-qc]}}.

\bibitem{Bl.al.10}
L.~Blanchet, S.~Detweiler, A.~{Le Tiec} and B.~F. Whiting, {\em Phys. Rev. D}
  {\bf 81}  (2010)   064004, \href{http://arxiv.org/abs/0910.0207}{{\ttfamily
  arXiv:0910.0207 [gr-qc]}}.

\bibitem{Bl.al2.10}
L.~Blanchet, S.~Detweiler, A.~{Le Tiec} and B.~F. Whiting, {\em Phys. Rev. D}
  {\bf 81}  (2010)   084033, \href{http://arxiv.org/abs/1002.0726}{{\ttfamily
  arXiv:1002.0726 [gr-qc]}}.

\bibitem{Ba.al.10}
L.~Barack, T.~Damour and N.~Sago, {\em Phys. Rev. D} {\bf 82}  (2010)   084036,
  \href{http://arxiv.org/abs/1008.0935}{{\ttfamily arXiv:1008.0935 [gr-qc]}}.

\bibitem{Sh.al.11}
A.~G. Shah, T.~S. Keidl, J.~L. Friedman, D.-H. Kim and L.~R. Price, {\em Phys.
  Rev. D} {\bf 83}  (2011)   064018,
  \href{http://arxiv.org/abs/1009.4876}{{\ttfamily arXiv:1009.4876 [gr-qc]}}.

\bibitem{BaSa.11}
L.~Barack and N.~Sago, {\em Phys. Rev. D} {\bf 83}  (2011)   084023,
  \href{http://arxiv.org/abs/1101.3331}{{\ttfamily arXiv:1101.3331 [gr-qc]}}.

\bibitem{Sh.al.14}
A.~G. Shah, J.~L. Friedman and B.~F. Whiting, {\em Phys. Rev. D} {\bf 89}
  (2014)   064042, \href{http://arxiv.org/abs/1312.1952}{{\ttfamily
  arXiv:1312.1952 [gr-qc]}}.

\bibitem{Do.al.14}
S.~R. Dolan, N.~Warburton, A.~I. Harte, A.~{Le Tiec}, B.~Wardell and L.~Barack,
  {\em Phys. Rev. D} {\bf 89}  (2014)   064011,
  \href{http://arxiv.org/abs/1312.0775}{{\ttfamily arXiv:1312.0775 [gr-qc]}}.

\bibitem{Do.al2.14}
S.~R. Dolan, P.~Nolan, A.~C. Ottewill, N.~Warburton and B.~Wardell  (2014)
  \href{http://arxiv.org/abs/1406.4890}{{\ttfamily arXiv:1406.4890 [gr-qc]}}.

\bibitem{BaSa.10}
L.~Barack and N.~Sago, {\em Phys. Rev. D} {\bf 81}  (2010)   084021,
  \href{http://arxiv.org/abs/1002.2386}{{\ttfamily arXiv:1002.2386 [gr-qc]}}.

\bibitem{Wa.al.12}
N.~Warburton, S.~Akcay, L.~Barack, J.~R. Gair and N.~Sago, {\em Phys. Rev. D}
  {\bf 85}  (2012)   061501(R),
  \href{http://arxiv.org/abs/1111.6908}{{\ttfamily arXiv:1111.6908 [gr-qc]}}.

\bibitem{Wa.al.14}
B.~Wardell, C.~R. Galley, A.~Zengino{\u{g}}lu, M.~Casals, S.~R. Dolan and A.~C.
  Ottewill, {\em Phys. Rev. D} {\bf 89}  (2014)   084021,
  \href{http://arxiv.org/abs/1401.1506}{{\ttfamily arXiv:1401.1506 [gr-qc]}}.

\bibitem{Sh.al.12}
A.~G. Shah, J.~L. Friedman and T.~S. Keidl, {\em Phys. Rev. D} {\bf 86}  (2012)
    084059, \href{http://arxiv.org/abs/1207.5595}{{\ttfamily arXiv:1207.5595
  [gr-qc]}}.

\bibitem{He.al.14}
A.~Heffernan, A.~Ottewill and B.~Wardell, {\em Phys. Rev. D} {\bf 89}  (2014)
  024030, \href{http://arxiv.org/abs/1211.6446}{{\ttfamily arXiv:1211.6446
  [gr-qc]}}.

\bibitem{Is.al.14}
S.~Isoyama {\em et~al.}  (2014)
  \href{http://arxiv.org/abs/1404.6133}{{\ttfamily arXiv:1404.6133 [gr-qc]}}.

\bibitem{De.12}
S.~Detweiler, {\em Phys. Rev. D} {\bf 85}  (2012)   044048,
  \href{http://arxiv.org/abs/1107.2098}{{\ttfamily arXiv:1107.2098 [gr-qc]}}.

\bibitem{Gr.12}
S.~E. Gralla, {\em Phys. Rev. D} {\bf 85}  (2012)   124011,
  \href{http://arxiv.org/abs/1203.3189}{{\ttfamily arXiv:1203.3189 [gr-qc]}}.

\bibitem{Po.12}
A.~Pound, {\em Phys. Rev. Lett.} {\bf 109}  (2012)   051101,
  \href{http://arxiv.org/abs/1201.5089}{{\ttfamily arXiv:1201.5089 [gr-qc]}}.

\bibitem{Po2.12}
A.~Pound, {\em Phys. Rev. D} {\bf 86}  (2012)   084019,
  \href{http://arxiv.org/abs/1206.6538}{{\ttfamily arXiv:1206.6538 [gr-qc]}}.

\bibitem{BuKa.13}
L.~M. Burko and G.~Khanna, {\em Phys. Rev. D} {\bf 88}  (2013)   024002,
  \href{http://arxiv.org/abs/1304.5296}{{\ttfamily arXiv:1304.5296 [gr-qc]}}.

\bibitem{PoMi.14}
A.~Pound and J.~Miller, {\em Phys. Rev. D} {\bf 89}  (2014)   104020,
  \href{http://arxiv.org/abs/1403.1843}{{\ttfamily arXiv:1403.1843 [gr-qc]}}.

\bibitem{Po.14}
A.~Pound  (2014) \href{http://arxiv.org/abs/1404.1543}{{\ttfamily
  arXiv:1404.1543 [gr-qc]}}.

\bibitem{SaTa.03}
M.~Sasaki and H.~Tagoshi, {\em Living Rev. Relativity} {\bf 6}  (2003)  ~5,
  \href{http://arxiv.org/abs/arXiv:gr-qc/0306120}{{\ttfamily
  arXiv:gr-qc/0306120}}.

\bibitem{Ba.09}
L.~Barack, {\em Class. Quant. Grav.} {\bf 26}  (2009)   213001,
  \href{http://arxiv.org/abs/0908.1664}{{\ttfamily arXiv:0908.1664 [gr-qc]}}.

\bibitem{De.11}
S.~Detweiler, { Elementary development of the gravitational self-force}, in
  {\em Mass and motion in general relativity\/},  eds. L.~Blanchet,
  A.~Spallicci and B.~Whiting, Fundamental Theories of Physics, Vol.~162
  (Springer, New York, 2011), p. 271.
\newblock \href{http://arxiv.org/abs/0908.4363}{{\ttfamily arXiv:0908.4363
  [gr-qc]}}.

\bibitem{Wa.11}
R.~M. Wald, { Introduction to gravitational self-force}, in {\em Mass and
  motion in general relativity\/},  eds. L.~Blanchet, A.~Spallicci and
  B.~Whiting, Fundamental Theories of Physics, Vol.~162 (Springer, New York,
  2011), p. 253.
\newblock \href{http://arxiv.org/abs/0907.0412}{{\ttfamily arXiv:0907.0412
  [gr-qc]}}.

\bibitem{Th.11}
J.~Thornburg, {\em GW Notes} {\bf 5}  (2011)  ~3,
  \href{http://arxiv.org/abs/1102.2857}{{\ttfamily arXiv:1102.2857 [gr-qc]}}.

\bibitem{Po.al.11}
E.~Poisson, A.~Pound and I.~Vega, {\em Living Rev. Relativity} {\bf 14}  (2011)
   ~7, \href{http://arxiv.org/abs/1102.0529}{{\ttfamily arXiv:1102.0529
  [gr-qc]}}.

\bibitem{Pr.05}
F.~Pretorius, {\em Phys. Rev. Lett.} {\bf 95}  (2005)   121101,
  \href{http://arxiv.org/abs/arXiv:gr-qc/0507014}{{\ttfamily
  arXiv:gr-qc/0507014}}.

\bibitem{Ca.al.06}
M.~Campanelli, C.~O. Lousto, P.~Marronetti and Y.~Zlochower, {\em Phys. Rev.
  Lett.} {\bf 96}  (2006)   111101,
  \href{http://arxiv.org/abs/arXiv:gr-qc/0511048}{{\ttfamily
  arXiv:gr-qc/0511048}}.

\bibitem{Ba.al.06}
J.~G. Baker, J.~Centrella, D.-I. Choi, M.~Koppitz and J.~{van Meter}, {\em
  Phys. Rev. Lett.} {\bf 96}  (2006)   111102,
  \href{http://arxiv.org/abs/arXiv:gr-qc/0511103}{{\ttfamily
  arXiv:gr-qc/0511103}}.

\bibitem{Ow.al.11}
R.~Owen {\em et~al.}, {\em Phys. Rev. Lett.} {\bf 106}  (2011)   151101,
  \href{http://arxiv.org/abs/1012.4869}{{\ttfamily arXiv:1012.4869 [gr-qc]}}.

\bibitem{Ba.al2.06}
J.~G. Baker, J.~Centrella, D.-I. Choi, M.~Koppitz and J.~{van Meter}, {\em
  Phys. Rev. D} {\bf 73}  (2006)   104002,
  \href{http://arxiv.org/abs/arXiv:gr-qc/0602026}{{\ttfamily
  arXiv:gr-qc/0602026}}.

\bibitem{Ba.al.07}
J.~G. Baker, J.~R. {van Meter}, S.~T. McWilliams, J.~Centrella and B.~J. Kelly,
  {\em Phys. Rev. Lett.} {\bf 99}  (2007)   181101,
  \href{http://arxiv.org/abs/arXiv:gr-qc/0612024}{{\ttfamily
  arXiv:gr-qc/0612024}}.

\bibitem{Ba.al3.07}
J.~G. Baker, M.~Campanelli, F.~Pretorius and Y.~Zlochower, {\em Class. Quant.
  Grav.} {\bf 24}  (2007)   S25,
  \href{http://arxiv.org/abs/arXiv:gr-qc/0701016}{{\ttfamily
  arXiv:gr-qc/0701016}}.

\bibitem{Bo.al2.07}
M.~Boyle {\em et~al.}, {\em Phys. Rev. D} {\bf 76}  (2007)   124038,
  \href{http://arxiv.org/abs/0710.0158}{{\ttfamily arXiv:0710.0158 [gr-qc]}}.

\bibitem{Bu.al.07}
A.~Buonanno, G.~B. Cook and F.~Pretorius, {\em Phys. Rev. D} {\bf 75}  (2007)
  124018, \href{http://arxiv.org/abs/arXiv:gr-qc/0610122}{{\ttfamily
  arXiv:gr-qc/0610122}}.

\bibitem{Ha.al.07}
M.~Hannam, S.~Husa, J.~A. Gonz{\'a}lez, U.~Sperhake and B.~Br{\"u}gmann, {\em
  Phys. Rev. D} {\bf 77}  (2007)   044020,
  \href{http://arxiv.org/abs/0706.1305}{{\ttfamily arXiv:0706.1305 [gr-qc]}}.

\bibitem{Sc.al.09}
M.~Scheel, M.~Boyle, T.~Chu, L.~E. Kidder, K.~D. Matthews and H.~P. Pfeiffer,
  {\em Phys. Rev. D} {\bf 79}  (2009)   024003,
  \href{http://arxiv.org/abs/0810.1767}{{\ttfamily arXiv:0810.1767 [gr-qc]}}.

\bibitem{Be.al.07}
E.~Berti {\em et~al.}, {\em Phys. Rev. D} {\bf 76}  (2007)   064034,
  \href{http://arxiv.org/abs/arXiv:gr-qc/0703053}{{\ttfamily
  arXiv:gr-qc/0703053}}.

\bibitem{Go.al.09}
J.~A. Gonz{\'a}lez, U.~Sperhake and B.~Br{\"u}gmann, {\em Phys. Rev. D} {\bf
  79}  (2009)   124006, \href{http://arxiv.org/abs/0811.3952}{{\ttfamily
  arXiv:0811.3952 [gr-qc]}}.

\bibitem{Sp.al.11}
U.~Sperhake, B.~Br{\"u}gmann, D.~M{\"u}ller and C.~F. Sopuerta, {\em Class.
  Quant. Grav.} {\bf 28}  (2011)   134004,
  \href{http://arxiv.org/abs/1012.3173}{{\ttfamily arXiv:1012.3173 [gr-qc]}}.

\bibitem{Bu.al.12}
L.~T. Buchman, H.~P. Pfeiffer, M.~A. Scheel and B.~Szil{\'a}gyi, {\em Phys.
  Rev. D} {\bf 86}  (2012)   084033,
  \href{http://arxiv.org/abs/1206.3015}{{\ttfamily arXiv:1206.3015 [gr-qc]}}.

\bibitem{Be.al.08}
E.~Berti, V.~Cardoso, J.~A. Gonz{\'a}lez, U.~Sperhake and B.~Br{\"u}gmann, {\em
  Class. Quant. Grav.} {\bf 25}  (2008)   114035,
  \href{http://arxiv.org/abs/0711.1097}{{\ttfamily arXiv:0711.1097 [gr-qc]}}.

\bibitem{Ha.al2.08}
M.~Hannam, S.~Husa, B.~Br{\"u}gmann and A.~Gopakumar, {\em Phys. Rev. D} {\bf
  78}  (2008)   104007, \href{http://arxiv.org/abs/0712.3787}{{\ttfamily
  arXiv:0712.3787 [gr-qc]}}.

\bibitem{Ch.al.09}
T.~Chu, H.~P. Pfeiffer and M.~A. Scheel, {\em Phys. Rev. D} {\bf 80}  (2009)
  124051, \href{http://arxiv.org/abs/0909.1313}{{\ttfamily arXiv:0909.1313
  [gr-qc]}}.

\bibitem{Ha.al.10}
M.~Hannam, S.~Husa, F.~Ohme, D.~M{\"u}ller and B.~Br{\"u}gmann, {\em Phys. Rev.
  D} {\bf 82}  (2010)   124008,
  \href{http://arxiv.org/abs/1007.4789}{{\ttfamily arXiv:1007.4789 [gr-qc]}}.

\bibitem{Ca.al.09}
M.~Campanelli, C.~O. Lousto, H.~Nakano and Y.~Zlochower, {\em Phys. Rev. D}
  {\bf 79}  (2009)   084010, \href{http://arxiv.org/abs/0808.0713}{{\ttfamily
  arXiv:0808.0713 [gr-qc]}}.

\bibitem{Sz.al.09}
B.~Szil{\'a}gyi, L.~Lindblom and M.~A. Scheel, {\em Phys. Rev. D} {\bf 80}
  (2009)   124010, \href{http://arxiv.org/abs/0909.3557}{{\ttfamily
  arXiv:0909.3557 [gr-qc]}}.

\bibitem{Sp.al.08}
U.~Sperhake, V.~Cardoso, F.~Pretorius, E.~Berti and J.~A. Gonz{\'a}lez, {\em
  Phys. Rev. Lett.} {\bf 101}  (2008)   161101,
  \href{http://arxiv.org/abs/0806.1738}{{\ttfamily arXiv:0806.1738 [gr-qc]}}.

\bibitem{Sp.al2.08}
U.~Sperhake, E.~Berti, V.~Cardoso, J.~A. Gonz{\'a}lez, B.~Br{\"u}gmann and
  M.~Ansorg, {\em Phys. Rev. D} {\bf 78}  (2008)   064069,
  \href{http://arxiv.org/abs/0710.3823}{{\ttfamily arXiv:0710.3823 [gr-qc]}}.

\bibitem{GoBr.10}
R.~Gold and B.~Br{\"u}gmann, {\em Class. Quant. Grav.} {\bf 27}  (2010)
  084035, \href{http://arxiv.org/abs/0911.3862}{{\ttfamily arXiv:0911.3862
  [gr-qc]}}.

\bibitem{Hi.al.10}
I.~Hinder, F.~Herrmann, P.~Laguna and D.~Shoemaker, {\em Phys. Rev. D} {\bf 82}
   (2010)   024033, \href{http://arxiv.org/abs/0806.1037}{{\ttfamily
  arXiv:0806.1037 [gr-qc]}}.

\bibitem{Sp.al2.11}
U.~Sperhake, V.~Cardoso, C.~D. Ott, E.~Schnetter and H.~Witek, {\em Phys. Rev.
  D} {\bf 84}  (2011)   084038,
  \href{http://arxiv.org/abs/1105.5391}{{\ttfamily arXiv:1105.5391 [gr-qc]}}.

\bibitem{GoBr.13}
R.~Gold and B.~Br{\"u}gmann, {\em Phys. Rev. D} {\bf 88}  (2013)   064051,
  \href{http://arxiv.org/abs/1209.4085}{{\ttfamily arXiv:1209.4085 [gr-qc]}}.

\bibitem{Da.al2.14}
T.~Damour, F.~Guercilena, I.~Hinder, S.~Hopper, A.~Nagar and L.~Rezzolla, {\em
  Phys. Rev. D} {\bf 89}  (2014)   081503(R),
  \href{http://arxiv.org/abs/1402.7307}{{\ttfamily arXiv:1402.7307 [gr-qc]}}.

\bibitem{Zl.al.12}
Y.~Zlochower, M.~Ponce and C.~O. Lousto, {\em Phys. Rev. D} {\bf 86}  (2012)
  104056, \href{http://arxiv.org/abs/1208.5494}{{\ttfamily arXiv:1208.5494
  [gr-qc]}}.

\bibitem{Hi.al2.13}
D.~Hilditch, S.~Bernuzzi, M.~Thierfelder, Z.~Cao, W.~Tichy and B.~Br{\"u}gmann,
  {\em Phys. Rev. D} {\bf 88}  (2013)   084057,
  \href{http://arxiv.org/abs/1212.2901}{{\ttfamily arXiv:1212.2901 [gr-qc]}}.

\bibitem{Mr.al.13}
A.~H. Mrou{\'e} {\em et~al.}, {\em Phys. Rev. Lett.} {\bf 111}  (2013)
  241104, \href{http://arxiv.org/abs/1304.6077}{{\ttfamily arXiv:1304.6077
  [gr-qc]}}.

\bibitem{Ha.al.14}
M.~Hannam {\em et~al.}  (2014) \href{http://arxiv.org/abs/1308.3271}{{\ttfamily
  arXiv:1308.3271 [gr-qc]}}.

\bibitem{Ha.14}
M.~Hannam, {\em Gen. Rel. Grav.} {\bf 46}  (2014)  ~1,
  \href{http://arxiv.org/abs/1312.3641}{{\ttfamily arXiv:1312.3641 [gr-qc]}}.

\bibitem{ReTi.12}
G.~Reifenberger and W.~Tichy, {\em Phys. Rev. D} {\bf 86}  (2012)   064003,
  \href{http://arxiv.org/abs/1205.5502}{{\ttfamily arXiv:1205.5502 [gr-qc]}}.

\bibitem{Ch.14}
T.~Chu, {\em Phys. Rev. D} {\bf 89}  (2014)   064062,
  \href{http://arxiv.org/abs/1310.7900}{{\ttfamily arXiv:1310.7900 [gr-qc]}}.

\bibitem{LoZl.11}
C.~O. Lousto and Y.~Zlochower, {\em Phys. Rev. Lett.} {\bf 106}  (2011)
  041101, \href{http://arxiv.org/abs/1009.0292}{{\ttfamily arXiv:1009.0292
  [gr-qc]}}.

\bibitem{Lo.al.12}
G.~Lovelace, M.~Boyle, M.~A. Scheel and B.~Szil{\'a}gyi, {\em Class. Quant.
  Grav.} {\bf 29}  (2012)   045003,
  \href{http://arxiv.org/abs/1110.2229}{{\ttfamily arXiv:1110.2229 [gr-qc]}}.

\bibitem{LoZl.13}
C.~O. Lousto and Y.~Zlochower, {\em Phys. Rev. D} {\bf 88}  (2013)   024001,
  \href{http://arxiv.org/abs/1304.3937}{{\ttfamily arXiv:1304.3937 [gr-qc]}}.

\bibitem{Ce.al.10}
J.~Centrella, J.~G. Baker, B.~J. Kelly and J.~R. {van Meter}, {\em Rev. Mod.
  Phys.} {\bf 82}  (2010)   3069,
  \href{http://arxiv.org/abs/1010.5260}{{\ttfamily arXiv:1010.5260 [gr-qc]}}.

\bibitem{Hi.10}
I.~Hinder, {\em Class. Quant. Grav.} {\bf 27}  (2010)   114004,
  \href{http://arxiv.org/abs/1001.5161}{{\ttfamily arXiv:1001.5161 [gr-qc]}}.

\bibitem{McW.11}
S.~T. McWilliams, {\em Class. Quant. Grav.} {\bf 28}  (2011)   134001,
  \href{http://arxiv.org/abs/1012.2872}{{\ttfamily arXiv:1012.2872 [gr-qc]}}.

\bibitem{Pf.12}
H.~P. Pfeiffer, {\em Class. Quant. Grav.} {\bf 29}  (2012)   124004,
  \href{http://arxiv.org/abs/1203.5166}{{\ttfamily arXiv:1203.5166 [gr-qc]}}.

\bibitem{Sp.al.13}
U.~Sperhake, E.~Berti and V.~Cardoso, {\em C. R. Phys.} {\bf 14}  (2013)   306,
  \href{http://arxiv.org/abs/1107.2819}{{\ttfamily arXiv:1107.2819 [gr-qc]}}.

\bibitem{Sz.14}
B.~Szil{\'a}gyi, {\em Int. J. Mod. Phys. D} {\bf 23}  (2014)   1430014,
  \href{http://arxiv.org/abs/1405.3693}{{\ttfamily arXiv:1405.3693 [gr-qc]}}.

\bibitem{BuDa.99}
A.~Buonanno and T.~Damour, {\em Phys. Rev. D} {\bf 59}  (1999)   084006,
  \href{http://arxiv.org/abs/arXiv:gr-qc/9811091}{{\ttfamily
  arXiv:gr-qc/9811091}}.

\bibitem{BuDa.00}
A.~Buonanno and T.~Damour, {\em Phys. Rev. D} {\bf 62}  (2000)   064015,
  \href{http://arxiv.org/abs/arXiv:gr-qc/0001013}{{\ttfamily
  arXiv:gr-qc/0001013}}.

\bibitem{Da.10}
T.~Damour, {\em Phys. Rev. D} {\bf 81}  (2010)   024017,
  \href{http://arxiv.org/abs/0910.5533}{{\ttfamily arXiv:0910.5533 [gr-qc]}}.

\bibitem{Ba.al.12}
E.~Barausse, A.~Buonanno and A.~{Le Tiec}, {\em Phys. Rev. D} {\bf 85}  (2012)
   064010, \href{http://arxiv.org/abs/1111.5610}{{\ttfamily arXiv:1111.5610
  [gr-qc]}}.

\bibitem{Ak.al.12}
S.~Akcay, L.~Barack, T.~Damour and N.~Sago, {\em Phys. Rev. D} {\bf 86}  (2012)
    104041, \href{http://arxiv.org/abs/1209.0964}{{\ttfamily arXiv:1209.0964
  [gr-qc]}}.

\bibitem{Da.01}
T.~Damour, {\em Phys. Rev. D} {\bf 64}  (2001)   124013,
  \href{http://arxiv.org/abs/arXiv:gr-qc/0103018}{{\ttfamily
  arXiv:gr-qc/0103018}}.

\bibitem{Da.al2.08}
T.~Damour, P.~Jaranowski and G.~Sch{\"a}fer, {\em Phys. Rev. D} {\bf 78}
  (2008)   024009, \href{http://arxiv.org/abs/0803.0915}{{\ttfamily
  arXiv:0803.0915 [gr-qc]}}.

\bibitem{BaBu.10}
E.~Barausse and A.~Buonanno, {\em Phys. Rev. D} {\bf 81}  (2010)   084024,
  \href{http://arxiv.org/abs/0912.3517}{{\ttfamily arXiv:0912.3517 [gr-qc]}}.

\bibitem{BaBu.11}
E.~Barausse and A.~Buonanno, {\em Phys. Rev. D} {\bf 84}  (2011)   104027,
  \href{http://arxiv.org/abs/1107.2904}{{\ttfamily arXiv:1107.2904 [gr-qc]}}.

\bibitem{Na.11}
A.~Nagar, {\em Phys. Rev. D} {\bf 84}  (2011)   084028,
  \href{http://arxiv.org/abs/1106.4349}{{\ttfamily arXiv:1106.4349 [gr-qc]}}.

\bibitem{BaJe.13}
S.~Balmelli and P.~Jetzer, {\em Phys. Rev D} {\bf 87}  (2013)   124036,
  \href{http://arxiv.org/abs/1305.5674}{{\ttfamily arXiv:1305.5674 [gr-qc]}}.

\bibitem{DaNa.08}
T.~Damour and A.~Nagar, {\em Phys. Rev. D} {\bf 77}  (2008)   024043,
  \href{http://arxiv.org/abs/0711.2628}{{\ttfamily arXiv:0711.2628 [gr-qc]}}.

\bibitem{Da.al.09}
T.~Damour, B.~R. Iyer and A.~Nagar, {\em Phys. Rev. D} {\bf 79}  (2009)
  064004, \href{http://arxiv.org/abs/0811.2069}{{\ttfamily arXiv:0811.2069
  [gr-qc]}}.

\bibitem{BiDa.12}
D.~Bini and T.~Damour, {\em Phys. Rev. D} {\bf 86}  (2012)   124012,
  \href{http://arxiv.org/abs/1210.2834}{{\ttfamily arXiv:1210.2834 [gr-qc]}}.

\bibitem{Da.al.97}
T.~Damour, B.~R. Iyer and B.~S. Sathyaprakash, {\em Phys. Rev. D} {\bf 57}
  (1997)   885, \href{http://arxiv.org/abs/arXiv:gr-qc/9708034}{{\ttfamily
  arXiv:gr-qc/9708034}}.

\bibitem{Da.al3.01}
T.~Damour, B.~R. Iyer and B.~S. Sathyaprakash, {\em Phys. Rev. D} {\bf 63}
  (2001)   044023, \href{http://arxiv.org/abs/arXiv:gr-qc/0010009}{{\ttfamily
  arXiv:gr-qc/0010009}}, \textit{{E}rratum:} Phys. Rev. D \textbf{72},
  029902(E) (2005).

\bibitem{Bl.02}
L.~Blanchet, {\em Phys. Rev. D} {\bf 65}  (2002)   124009,
  \href{http://arxiv.org/abs/arXiv:gr-qc/0112056}{{\ttfamily
  arXiv:gr-qc/0112056}}.

\bibitem{Bo.al.08}
M.~Boyle {\em et~al.}, {\em Phys. Rev. D} {\bf 78}  (2008)   104020,
  \href{http://arxiv.org/abs/0804.4184}{{\ttfamily arXiv:0804.4184 [gr-qc]}}.

\bibitem{Mr.al.08}
A.~H. Mrou{\'e}, L.~E. Kidder and S.~A. Teukolsky, {\em Phys. Rev. D} {\bf 78}
  (2008)   044004, \href{http://arxiv.org/abs/0805.2390}{{\ttfamily
  arXiv:0805.2390 [gr-qc]}}.

\bibitem{Fa.11}
M.~Favata, {\em Phys. Rev. D} {\bf 83}  (2011)   024027,
  \href{http://arxiv.org/abs/1008.4622}{{\ttfamily arXiv:1008.4622 [gr-qc]}}.

\bibitem{Pa.al.11}
Y.~Pan {\em et~al.}, {\em Phys. Rev. D} {\bf 84}  (2011)   124052,
  \href{http://arxiv.org/abs/1106.1021}{{\ttfamily arXiv:1106.1021 [gr-qc]}}.

\bibitem{Ta.al.12}
A.~Taracchini {\em et~al.}, {\em Phys. Rev. D} {\bf 86}  (2012)   024011,
  \href{http://arxiv.org/abs/1202.0790}{{\ttfamily arXiv:1202.0790 [gr-qc]}}.

\bibitem{Pa.al.14}
Y.~Pan {\em et~al.}, {\em Phys. Rev. D} {\bf 89}  (2014)   084006,
  \href{http://arxiv.org/abs/1307.6232}{{\ttfamily arXiv:1307.6232 [gr-qc]}}.

\bibitem{Pa.al2.14}
Y.~Pan {\em et~al.}, {\em Phys. Rev. D} {\bf 89}  (2014)   061501(R),
  \href{http://arxiv.org/abs/1311.2565}{{\ttfamily arXiv:1311.2565 [gr-qc]}}.

\bibitem{Ta.al.14}
A.~Taracchini {\em et~al.}, {\em Phys. Rev. D} {\bf 89}  (2014)   061502(R),
  \href{http://arxiv.org/abs/1311.2544}{{\ttfamily arXiv:1311.2544 [gr-qc]}}.

\bibitem{Da.al.13}
T.~Damour, A.~Nagar and S.~Bernuzzi, {\em Phys. Rev. D} {\bf 87}  (2013)
  084035, \href{http://arxiv.org/abs/1212.4357}{{\ttfamily arXiv:1212.4357
  [gr-qc]}}.

\bibitem{DaNa.14}
T.~Damour and A.~Nagar, {\em Phys. Rev. D} {\bf 90}  (2014)   044018,
  \href{http://arxiv.org/abs/1406.6913}{{\ttfamily arXiv:1406.6913 [gr-qc]}}.

\bibitem{DaNa.11}
T.~Damour and A.~Nagar, { The effective one-body description of the two-body
  problem}, in {\em Mass and motion in general relativity\/},  eds.
  L.~Blanchet, A.~Spallicci and B.~Whiting, Fundamental Theories of Physics,
  Vol.~162 (Springer, New York, 2011), p. 211.
\newblock \href{http://arxiv.org/abs/0906.1769}{{\ttfamily arXiv:0906.1769
  [gr-qc]}}.

\bibitem{Da.14}
T.~Damour, { The general relativistic two body problem and the effective one
  body formalism}, in {\em General relativity, cosmology and astrophysics:
  Perspectives 100 years after {E}instein's stay in {P}rague\/},  eds.
  J.~Bi\v{c}{\'a}k and T.~Ledvinka, Fundamental Theories of Physics, Vol.~177
  (Springer, New York, 2014), p. 111.
\newblock \href{http://arxiv.org/abs/1212.3169}{{\ttfamily arXiv:1212.3169
  [gr-qc]}}.

\bibitem{Da2.14}
T.~Damour, { The general relativistic two body problem}, in {\em Frontiers in
  relativistic celestial mechanics. {V}olume 1: {T}heory\/},  ed. S.~M.
  Kopeikin, De Gruyter studies in mathematical physics, Vol.~21 (de Gruyter,
  Berlin, 2014).
\newblock \href{http://arxiv.org/abs/1312.3505}{{\ttfamily arXiv:1312.3505
  [gr-qc]}}.

\bibitem{Pa.al.08}
Y.~Pan {\em et~al.}, {\em Phys. Rev. D} {\bf 77}  (2008)   024014,
  \href{http://arxiv.org/abs/0704.1964}{{\ttfamily arXiv:0704.1964 [gr-qc]}}.

\bibitem{Lo.al2.10}
C.~O. Lousto, H.~Nakano, Y.~Zlochower and M.~Campanelli, {\em Phys. Rev. Lett.}
  {\bf 104}  (2010)   211101, \href{http://arxiv.org/abs/1001.2316}{{\ttfamily
  arXiv:1001.2316 [gr-qc]}}.

\bibitem{Na.al.11}
H.~Nakano, Y.~Zlochower, C.~O. Lousto and M.~Campanelli, {\em Phys. Rev. D}
  {\bf 84}  (2011)   124006, \href{http://arxiv.org/abs/1108.4421}{{\ttfamily
  arXiv:1108.4421 [gr-qc]}}.

\bibitem{Na.13}
A.~Nagar, {\em Phys. Rev. D} {\bf 88}  (2013)   121501(R),
  \href{http://arxiv.org/abs/1306.6299}{{\ttfamily arXiv:1306.6299 [gr-qc]}}.

\bibitem{Hi.al.14}
I.~Hinder {\em et~al.}, {\em Class. Quant. Grav.} {\bf 31}  (2014)   025012,
  \href{http://arxiv.org/abs/1307.5307}{{\ttfamily arXiv:1307.5307 [gr-qc]}}.

\bibitem{Mr.al.10}
A.~H. Mrou{\'e}, H.~P. Pfeiffer, L.~E. Kidder and S.~A. Teukolsky, {\em Phys.
  Rev. D} {\bf 82}  (2010)   124016,
  \href{http://arxiv.org/abs/1004.4697}{{\ttfamily arXiv:1004.4697 [gr-qc]}}.

\bibitem{Le.al.11}
A.~{Le Tiec} {\em et~al.}, {\em Phys. Rev. Lett.} {\bf 107}  (2011)   141101,
  \href{http://arxiv.org/abs/1106.3278}{{\ttfamily arXiv:1106.3278 [gr-qc]}}.

\bibitem{Da.al.12}
T.~Damour, A.~Nagar, D.~Pollney and C.~Reisswig, {\em Phys. Rev. Lett.} {\bf
  108}  (2012)   131101, \href{http://arxiv.org/abs/1110.2938}{{\ttfamily
  arXiv:1110.2938 [gr-qc]}}.

\bibitem{Le.al2.12}
A.~{Le Tiec}, E.~Barausse and A.~Buonanno, {\em Phys. Rev. Lett.} {\bf 108}
  (2012)   131103, \href{http://arxiv.org/abs/1111.5609}{{\ttfamily
  arXiv:1111.5609 [gr-qc]}}.

\bibitem{Hi.al.13}
T.~Hinderer {\em et~al.}, {\em Phys. Rev. D} {\bf 88}  (2013)   084005,
  \href{http://arxiv.org/abs/1309.0544}{{\ttfamily arXiv:1309.0544 [gr-qc]}}.

\bibitem{Le.al.13}
A.~{Le Tiec} {\em et~al.}, {\em Phys. Rev. D} {\bf 88}  (2013)   124027,
  \href{http://arxiv.org/abs/1309.0541}{{\ttfamily arXiv:1309.0541 [gr-qc]}}.

\bibitem{BiDa.13}
D.~Bini and T.~Damour, {\em Phys. Rev. D} {\bf 87}  (2013)   121501(R),
  \href{http://arxiv.org/abs/1305.4884}{{\ttfamily arXiv:1305.4884 [gr-qc]}}.

\bibitem{BiDa.14}
D.~Bini and T.~Damour, {\em Phys. Rev. D} {\bf 89}  (2014)   064063,
  \href{http://arxiv.org/abs/1312.2503}{{\ttfamily arXiv:1312.2503 [gr-qc]}}.

\bibitem{BiDa2.14}
D.~Bini and T.~Damour, {\em Phys. Rev. D} {\bf 89}  (2014)   104047,
  \href{http://arxiv.org/abs/1403.2366}{{\ttfamily arXiv:1403.2366 [gr-qc]}}.

\bibitem{Bl.al.14}
L.~Blanchet, G.~Faye and B.~F. Whiting, {\em Phys. Rev. D} {\bf 89}  (2014)
  064026, \href{http://arxiv.org/abs/1312.2975}{{\ttfamily arXiv:1312.2975
  [gr-qc]}}.

\bibitem{Bl.al2.14}
L.~Blanchet, G.~Faye and B.~F. Whiting, {\em Phys. Rev. D} {\bf 90}  (2014)
  044017, \href{http://arxiv.org/abs/1405.5151}{{\ttfamily arXiv:1405.5151
  [gr-qc]}}.

\bibitem{BiDa3.14}
D.~Bini and T.~Damour, {\em Phys. Rev. D} {\bf 90}  (2014)   024039,
  \href{http://arxiv.org/abs/1404.2747}{{\ttfamily arXiv:1404.2747 [gr-qc]}}.

\bibitem{Fa.al.12}
G.~Faye, S.~Marsat, L.~Blanchet and B.~R. Iyer, {\em Class. Quant. Grav.} {\bf
  29}  (2012)   175004, \href{http://arxiv.org/abs/1204.1043}{{\ttfamily
  arXiv:1204.1043 [gr-qc]}}.

\bibitem{Da.al.71}
M.~Davis, R.~Ruffini, W.~H. Press and R.~H. Price, {\em Phys. Rev. Lett.} {\bf
  27}  (1971)   1466.

\bibitem{De.79}
S.~L. Detweiler, { Black holes and gravitational waves: Perturbation analysis},
  in {\em Sources of gravitational radiation\/},  ed. L.~Smarr (Cambridge
  University Press, Cambridge, 1979), p. 211.

\bibitem{Sm.79}
L.~Smarr, { Gauge conditions, radiation formulae and the two black hole
  collision}, in {\em Sources of gravitational radiation\/},  ed. L.~Smarr
  (Cambridge University Press, Cambridge, 1979), p. 245.

\bibitem{An.al.95}
P.~Anninos, D.~Hobill, E.~Seidel, L.~Smarr and W.-M. Suen, {\em Phys. Rev. D}
  {\bf 52}  (1995)   2044.

\bibitem{Is1.68}
R.~A. Isaacson, {\em Phys. Rev.} {\bf 166}  (1968)   1263.

\bibitem{Is2.68}
R.~A. Isaacson, {\em Phys. Rev.} {\bf 166}  (1968)   1272.

\bibitem{Ma.al.96}
S.~Mano, H.~Suzuki and E.~Takasugi, {\em Prog. Theor. Phys.} {\bf 95}  (1996)
  1079, \href{http://arxiv.org/abs/arXiv:gr-qc/9603020}{{\ttfamily
  arXiv:gr-qc/9603020}}.

\bibitem{MaTa.97}
S.~Mano and E.~Takasugi, {\em Prog. Theor. Phys.} {\bf 97}  (1997)   213,
  \href{http://arxiv.org/abs/arXiv:gr-qc/9611014}{{\ttfamily
  arXiv:gr-qc/9611014}}.

\bibitem{Ta.al.93}
T.~Tanaka, M.~Shibata, M.~Sasaki, H.~Tagoshi and T.~Nakamura, {\em Prog. Theor.
  Phys.} {\bf 90}  (1993)  ~65.

\bibitem{TaSa.94}
H.~Tagoshi and M.~Sasaki, {\em Prog. Theor. Phys.} {\bf 92}  (1994)   745,
  \href{http://arxiv.org/abs/arXiv:gr-qc/9405062}{{\ttfamily
  arXiv:gr-qc/9405062}}.

\bibitem{Ta.al2.96}
T.~Tanaka, H.~Tagoshi and M.~Sasaki, {\em Prog. Theor. Phys.} {\bf 96}  (1996)
   1087, \href{http://arxiv.org/abs/arXiv:gr-qc/9701050}{{\ttfamily
  arXiv:gr-qc/9701050}}.

\bibitem{Mi.al2.97}
Y.~Mino, M.~Sasaki, M.~Shibata, H.~Tagoshi and T.~Tanaka, {\em Prog. Theor.
  Phys. Suppl.} {\bf 128}  (1997)  ~1,
  \href{http://arxiv.org/abs/arXiv:gr-qc/9712057}{{\ttfamily
  arXiv:gr-qc/9712057}}.

\bibitem{Fu.12}
R.~Fujita, {\em Prog. Theor. Phys.} {\bf 127}  (2012)   583,
  \href{http://arxiv.org/abs/1104.5615}{{\ttfamily arXiv:1104.5615 [gr-qc]}}.

\bibitem{Fu2.12}
R.~Fujita, {\em Prog. Theor. Phys.} {\bf 128}  (2012)   971,
  \href{http://arxiv.org/abs/1211.5535}{{\ttfamily arXiv:1211.5535 [gr-qc]}}.

\bibitem{Cu.al2.93}
C.~Cutler, L.~S. Finn, E.~Poisson and G.~J. Sussman, {\em Phys. Rev. D} {\bf
  47}  (1993)   1511.

\bibitem{TaNa.94}
H.~Tagoshi and T.~Nakamura, {\em Phys. Rev. D} {\bf 49}  (1994)   4016.

\bibitem{FuTa.04}
R.~Fujita and H.~Tagoshi, {\em Prog. Theor. Phys.} {\bf 112}  (2004)   415,
  \href{http://arxiv.org/abs/arXiv:gr-qc/0410018}{{\ttfamily
  arXiv:gr-qc/0410018}}.

\bibitem{Ki.95}
L.~E. Kidder, {\em Phys. Rev. D} {\bf 52}  (1995)   821,
  \href{http://arxiv.org/abs/arXiv:gr-qc/9506022}{{\ttfamily
  arXiv:gr-qc/9506022}}.

\bibitem{Bl.al.06}
L.~Blanchet, A.~Buonanno and G.~Faye, {\em Phys. Rev. D} {\bf 74}  (2006)
  104034, \href{http://arxiv.org/abs/arXiv:gr-qc/0605140}{{\ttfamily
  arXiv:gr-qc/0605140}}, \textit{{E}rrata:} Phys. Rev. D \textbf{75}, 049903(E)
  (2007) \& Phys. Rev. D \textbf{81}, 089901(E) (2010).

\bibitem{Po2.93}
E.~Poisson, {\em Phys. Rev. D} {\bf 48}  (1993)   1860.

\bibitem{Sh.al.95}
M.~Shibata, M.~Sasaki, H.~Tagoshi and T.~Tanaka, {\em Phys. Rev. D} {\bf 51}
  (1995)   1646, \href{http://arxiv.org/abs/arXiv:gr-qc/9409054}{{\ttfamily
  arXiv:gr-qc/9409054}}.

\bibitem{Ta.al.96}
H.~Tagoshi, M.~Shibata, T.~Tanaka and M.~Sasaki, {\em Phys. Rev. D} {\bf 54}
  (1996)   1439, \href{http://arxiv.org/abs/arXiv:gr-qc/9603028}{{\ttfamily
  arXiv:gr-qc/9603028}}.

\bibitem{Sh.14}
A.~G. Shah, {\em Phys. Rev. D} {\bf 90}  (2014)   044025,
  \href{http://arxiv.org/abs/1403.2697}{{\ttfamily arXiv:1403.2697 [gr-qc]}}.

\bibitem{Ru.al.08}
M.~Ruiz, M.~Alcubierre, D.~N{\'u}{\~n}ez and R.~Takahashi, {\em Gen. Rel.
  Grav.} {\bf 40}  (2008)   1705, \href{http://arxiv.org/abs/v}{{\ttfamily
  arXiv:v [gr-qc]}}.

\bibitem{Me.al.04}
D.~Meritt, M.~Milosavljevi{\'c}, M.~Favata, S.~A. Hugues and D.~E. Holz, {\em
  Astrophys. J.} {\bf 607}  (2004)  ~L9,
  \href{http://arxiv.org/abs/arXiv:astro-ph/0402057}{{\ttfamily
  arXiv:astro-ph/0402057}}.

\bibitem{Fi.83}
M.~J. Fitchett, {\em Mon. Not. Roy. Astron. Soc.} {\bf 203}  (1983)   1049.

\bibitem{FiDe.84}
M.~J. Fitchett and S.~Detweiler, {\em Mon. Not. Roy. Astron. Soc.} {\bf 211}
  (1984)   933.

\bibitem{Bl.al.05}
L.~Blanchet, M.~S.~S. Qusailah and C.~M. Will, {\em Astrophys. J.} {\bf 635}
  (2005)   508, \href{http://arxiv.org/abs/arXiv:astro-ph/0507692}{{\ttfamily
  arXiv:astro-ph/0507692}}.

\bibitem{So.al.06}
C.~F. Sopuerta, N.~Yunes and P.~Laguna, {\em Phys. Rev. D} {\bf 74}  (2006)
  124010, \href{http://arxiv.org/abs/arXiv:astro-ph/0608600}{{\ttfamily
  arXiv:astro-ph/0608600}}, \textit{{E}rrata:} Phys. Rev. D \textbf{75},
  069903(E) (2007) \& Phys. Rev. D \textbf{78}, 049901(E) (2008).

\bibitem{DaGo.06}
T.~Damour and A.~Gopakumar, {\em Phys. Rev. D} {\bf 73}  (2006)   124006,
  \href{http://arxiv.org/abs/arXiv:gr-qc/0602117}{{\ttfamily
  arXiv:gr-qc/0602117}}.

\bibitem{LeBl.10}
A.~{Le Tiec} and L.~Blanchet, {\em Class. Quant. Grav.} {\bf 27}  (2010)
  045008, \href{http://arxiv.org/abs/0910.4593}{{\ttfamily arXiv:0910.4593
  [gr-qc]}}.

\bibitem{Le.al.10}
A.~{Le Tiec}, L.~Blanchet and C.~M. Will, {\em Class. Quant. Grav.} {\bf 27}
  (2010)   121001, \href{http://arxiv.org/abs/0910.4594}{{\ttfamily
  arXiv:0910.4594 [gr-qc]}}.

\bibitem{Re.al.10}
L.~Rezzolla, R.~P. Macedo and J.~L. Jaramillo, {\em Phys. Rev. Lett.} {\bf 104}
   (2010)   221101, \href{http://arxiv.org/abs/1003.0873}{{\ttfamily
  arXiv:1003.0873 [gr-qc]}}.

\bibitem{Mi.al.12}
C.~K. Mishra, K.~G. Arun and B.~R. Iyer, {\em Phys. Rev. D} {\bf 85}  (2012)
  044021, \href{http://arxiv.org/abs/1111.2701}{{\ttfamily arXiv:1111.2701
  [gr-qc]}}.

\bibitem{Mi.12}
C.~K. Mishra, {\em Phys. Rev. D} {\bf 85}  (2012)   104046,
  \href{http://arxiv.org/abs/1202.6213}{{\ttfamily arXiv:1202.6213 [gr-qc]}}.

\bibitem{Ba.al3.06}
J.~G. Baker {\em et~al.}, {\em Astrophys. J.} {\bf 653}  (2006)   L93,
  \href{http://arxiv.org/abs/arXiv:astro-ph/0603204}{{\ttfamily
  arXiv:astro-ph/0603204}}.

\bibitem{Go.al.07}
J.~A. Gonz{\'a}lez, U.~Sperhake, B.~Br{\"u}gmann, M.~Hannam and S.~Husa, {\em
  Phys. Rev. Lett.} {\bf 98}  (2007)   091101,
  \href{http://arxiv.org/abs/arXiv:gr-qc/0610154}{{\ttfamily
  arXiv:gr-qc/0610154}}.

\bibitem{He.al.07}
F.~Herrmann, I.~Hinder, D.~Shoemaker and P.~Laguna, {\em Class. Quant. Grav.}
  {\bf 24}  (2007)   S33,
  \href{http://arxiv.org/abs/arXiv:gr-qc/0601026}{{\ttfamily
  arXiv:gr-qc/0601026}}.

\bibitem{Ko.al.07}
M.~Koppitz {\em et~al.}, {\em Phys. Rev. Lett.} {\bf 99}  (2007)   041102,
  \href{http://arxiv.org/abs/arXiv:gr-qc/0701163}{{\ttfamily
  arXiv:gr-qc/0701163}}.

\bibitem{Ja.al.12}
J.~L. Jaramillo, R.~P. Macedo, P.~Moesta and L.~Rezzolla, {\em Phys. Rev. D}
  {\bf 85}  (2012)   084030, \href{http://arxiv.org/abs/1108.0060}{{\ttfamily
  arXiv:1108.0060 [gr-qc]}}.

\bibitem{Le.al.12}
A.~{Le Tiec}, L.~Blanchet and B.~F. Whiting, {\em Phys. Rev. D} {\bf 85}
  (2012)   064039, \href{http://arxiv.org/abs/1111.5378}{{\ttfamily
  arXiv:1111.5378 [gr-qc]}}.

\bibitem{Bl.al.11}
L.~Blanchet, S.~Detweiler, A.~{Le Tiec} and B.~F. Whiting, { High-accuracy
  comparison between the post-{N}ewtonian and self-force dynamics of black-hole
  binaries}, in {\em Mass and motion in general relativity\/},  eds.
  L.~Blanchet, A.~Spallicci and B.~Whiting, Fundamental Theories of Physics,
  Vol.~162 (Springer, New York, 2011), p. 415.
\newblock \href{http://arxiv.org/abs/1007.2614}{{\ttfamily arXiv:1007.2614
  [gr-qc]}}.

\bibitem{Ba.al.73}
J.~M. Bardeen, B.~Carter and S.~W. Hawking, {\em Commun. Math. Phys.} {\bf 31}
  (1973)   161.

\bibitem{Fr.al.02}
J.~L. Friedman, K.~Ury\={u} and M.~Shibata, {\em Phys. Rev. D} {\bf 65}  (2002)
    064035, \href{http://arxiv.org/abs/arXiv:gr-qc/0108070}{{\ttfamily
  arXiv:gr-qc/0108070}}, \textit{{E}rratum:} Phys. Rev. D \textbf{70},
  129904(E) (2004).

\bibitem{GrLe.13}
S.~E. Gralla and A.~{Le Tiec}, {\em Phys. Rev. D} {\bf 88}  (2013)   044021,
  \href{http://arxiv.org/abs/1210.8444}{{\ttfamily arXiv:1210.8444 [gr-qc]}}.

\bibitem{deS.16}
W.~{de Sitter}, {\em Mon. Not. Roy. Astron. Soc.} {\bf 77}  (1916)   155.

\bibitem{Str}
N.~Straumann, {\em General relativity}, second edn. (Springer, New York, 2013).

\bibitem{Wa.al.13}
N.~Warburton, L.~Barack and N.~Sago, {\em Phys. Rev. D} {\bf 87}  (2013)
  084012, \href{http://arxiv.org/abs/1301.3918}{{\ttfamily arXiv:1301.3918
  [gr-qc]}}.

\bibitem{DaSc.88}
T.~Damour and G.~Sch{\"a}fer, {\em Nuovo Cim. B} {\bf 101}  (1988)   127.

\bibitem{Me.al2.04}
R.-M. Memmesheimer, A.~Gopakumar and G.~Sch{\"a}fer, {\em Phys. Rev. D} {\bf
  70}  (2004)   104011,
  \href{http://arxiv.org/abs/arXiv:gr-qc/0407049}{{\ttfamily
  arXiv:gr-qc/0407049}}.

\bibitem{Ro.38}
H.~P. Robertson, {\em Ann. Math.} {\bf 39}  (1938)   101.

\bibitem{Da.al.00}
T.~Damour, P.~Jaranowski and G.~Sch{\"a}fer, {\em Phys. Rev. D} {\bf 62}
  (2000)   044024, \href{http://arxiv.org/abs/arXiv:gr-qc/9912092}{{\ttfamily
  arXiv:gr-qc/9912092}}.

\bibitem{Cu.al.94}
C.~Cutler, D.~Kennefick and E.~Poisson, {\em Phys. Rev. D} {\bf 50}  (1994)
  3816.

\bibitem{Da.al3.00}
T.~Damour, P.~Jaranowski and G.~Sch{\"a}fer, {\em Phys. Rev. D} {\bf 62}
  (2000)   084011, \href{http://arxiv.org/abs/arXiv:gr-qc/0005034}{{\ttfamily
  arXiv:gr-qc/0005034}}.

\bibitem{BlFa2.00}
L.~Blanchet and G.~Faye, {\em Phys. Lett. A} {\bf 271}  (2000)  ~58,
  \href{http://arxiv.org/abs/arXiv:gr-qc/0004009}{{\ttfamily
  arXiv:gr-qc/0004009}}.

\bibitem{JaSc.12}
P.~Jaranowski and G.~Sch{\"a}fer, {\em Phys. Rev. D} {\bf 86}  (2012)
  061503(R), \href{http://arxiv.org/abs/1207.5448}{{\ttfamily arXiv:1207.5448
  [gr-qc]}}.

\bibitem{AsMa.79}
A.~Ashtekar and A.~Magnon-Ashtekar, {\em Phys. Rev. Lett.} {\bf 43}  (1979)
  181, \textit{{E}rratum:} Phys. Rev. Lett. \textbf{43}, 649 (1979).

\bibitem{Po.al3.11}
D.~Pollney, C.~Reisswig, E.~Schnetter, N.~Dorband and P.~Diener, {\em Phys.
  Rev. D} {\bf 83}  (2011)   044045,
  \href{http://arxiv.org/abs/0910.3803}{{\ttfamily arXiv:0910.3803 [gr-qc]}}.

\bibitem{Aj.al.07}
P.~Ajith {\em et~al.}, {\em Class. Quant. Grav.} {\bf 24}  (2007)   S689,
  \href{http://arxiv.org/abs/0704.3764}{{\ttfamily arXiv:0704.3764 [gr-qc]}}.

\bibitem{Aj.al.08}
P.~Ajith {\em et~al.}, {\em Phys. Rev. D} {\bf 77}  (2008)   104017,
  \href{http://arxiv.org/abs/0710.2335}{{\ttfamily arXiv:0710.2335 [gr-qc]}},
  \textit{{E}rratum:} Phys. Rev. D \textbf{79}, 129901(E) (2009).

\bibitem{Sa.al.10}
L.~Santamar{\'i}a {\em et~al.}, {\em Phys. Rev. D} {\bf 82}  (2010)   064016,
  \href{http://arxiv.org/abs/1005.3306}{{\ttfamily arXiv:1005.3306 [gr-qc]}}.

\bibitem{Aj.al.11}
P.~Ajith {\em et~al.}, {\em Phys. Rev. Lett.} {\bf 106}  (2011)   241101,
  \href{http://arxiv.org/abs/0909.2867}{{\ttfamily arXiv:0909.2867 [gr-qc]}}.

\bibitem{Ha.al2.10}
M.~Hannam, S.~Husa, F.~Ohme and P.~Ajith, {\em Phys. Rev. D} {\bf 82}  (2010)
  124052, \href{http://arxiv.org/abs/1008.2961}{{\ttfamily arXiv:1008.2961
  [gr-qc]}}.

\bibitem{Bo.11}
M.~Boyle, {\em Phys. Rev. D} {\bf 84}  (2011)   064013,
  \href{http://arxiv.org/abs/1103.5088}{{\ttfamily arXiv:1103.5088 [gr-qc]}}.

\bibitem{Da.al.11}
T.~Damour, A.~Nagar and M.~Trias, {\em Phys. Rev. D} {\bf 83}  (2011)   024006,
  \href{http://arxiv.org/abs/1009.5998}{{\ttfamily arXiv:1009.5998 [gr-qc]}}.

\bibitem{Ma.al.11}
I.~MacDonald, S.~Nissanke and H.~P. Pfeiffer, {\em Class. Quant. Grav.} {\bf
  28}  (2011)   134002, \href{http://arxiv.org/abs/1102.5128}{{\ttfamily
  arXiv:1102.5128 [gr-qc]}}.

\bibitem{Ma.al2.13}
I.~MacDonald {\em et~al.}, {\em Phys. Rev. D} {\bf 87}  (2013)   024009,
  \href{http://arxiv.org/abs/1210.3007}{{\ttfamily arXiv:1210.3007 [gr-qc]}}.

\bibitem{Oh.12}
F.~Ohme, {\em Class. Quant. Grav.} {\bf 29}  (2012)   124002,
  \href{http://arxiv.org/abs/1111.3737}{{\ttfamily arXiv:1111.3737 [gr-qc]}}.

\bibitem{Am.al.07}
P.~Amaro-Seoane {\em et~al.}, {\em Class. Quant. Grav.} {\bf 24}  (2007)
  R113, \href{http://arxiv.org/abs/arXiv:astro-ph/0703495}{{\ttfamily
  arXiv:astro-ph/0703495}}.

\bibitem{Sa.al.12}
B.~Sathyaprakash {\em et~al.}, {\em Class. Quant. Grav.} {\bf 29}  (2012)
  124013, \href{http://arxiv.org/abs/1206.0331}{{\ttfamily arXiv:1206.0331
  [gr-qc]}}.

\bibitem{BiPa.13}
M.-A. Bizouard and M.~A. Papa, {\em C. R. Phys.} {\bf 14}  (2013)   352,
  \href{http://arxiv.org/abs/1304.4984}{{\ttfamily arXiv:1304.4984 [gr-qc]}}.

\bibitem{Lo.al.08}
G.~Lovelace, R.~Owen, H.~P. Pfeiffer and T.~Chu, {\em Phys. Rev. D} {\bf 78}
  (2008)   084017, \href{http://arxiv.org/abs/0805.4192}{{\ttfamily
  arXiv:0805.4192 [gr-qc]}}.

\bibitem{StPu.10}
J.~Steinhoff and D.~Puetzfeld, {\em Phys. Rev. D} {\bf 81}  (2010)   044019,
  \href{http://arxiv.org/abs/0909.3756}{{\ttfamily arXiv:0909.3756 [gr-qc]}}.

\end{thebibliography}

\end{document}